\documentclass[twocolumn,10pt]{IEEEtran}
\def\BibTeX{{\rm B\kern-.05em{\sc i\kern-.025em b}\kern-.08em T\kern-.1667em\lower.7ex\hbox{E}\kern-.125emX}}

\begin{document}

\vspace{-0.5cm}
\title{Randomized Distributed Configuration Management of Wireless Networks: Multi-layer Markov Random Fields and Near-Optimality}

{\small
\author{Sung-eok Jeon and Chuanyi Ji, Senior Member, IEEE
\thanks{S. Jeon and C. Ji are with the School of Electrical and Computer Engineering, Georgia Institute of Technology, Atlanta, GA 30332 USA.
E-mail: sujeon@microsoft.com and jic@ece.gatech.edu}
}
}

\maketitle
\vspace{-1.0cm}

%%%%%%%%%%%%%%%%%%
\begin{abstract}

Distributed configuration management is imperative for wireless infrastructureless networks where each node adjusts locally its physical and logical configuration through information exchange with neighbors. Many algorithms have been developed with promising results in this area. However, two issues remain open. The first is the optimality, i.e., whether a distributed algorithm results in a near-optimal {\it network} configuration. The second is the complexity, i.e., whether a distributed  algorithm scales gracefully with network size ($N$).  We study these issues through modeling, analysis, and randomized distributed algorithms. 

Modeling defines the optimality. We first derive a global probabilistic model for a network configuration which characterizes jointly the statistical spatial dependence of a physical- and a logical-configuration. The model is a Gibbs distribution that results from internal network properties on node positions, wireless channels and interference; and external management constraints on physical connectivity, signal quality and configuration costs. We then show that a local model which approximates the global model is a two-layer Markov Random Field or a random bond model. The complexity of the local model is the communication range among nodes. The local model is near-optimal when the approximation error to the global model is within a given error bound. 
\textcolor{black}{
We analyze the trade-off between an approximation error and complexity, and derive sufficient conditions on the near-optimality of the local model. We show that when the power attenuation of a wireless channel is larger  than $4$, a node needs to communicate with more than $O(1)$ neighbors for a local model to be near optimal. For a slowly decaying channel with power attenuation less than 4, a node may need to communicate with more than $O(\sqrt[3]{N})$ neighbors to result in a bounded approximation error. 
}  
The two-layer Markov Random Fields enable a class of randomized distributed algorithms that allow a node to self-configure based on information from its neighbors. The distributed algorithms are applied to examples of (a) forming a 1-connected physical topology, (b) configuring a logical topology that maximizes the spatial channel reuse, and (c)  reconfiguring from failures, both sequentially and jointly. We validate the model, the analysis and the randomized distributed algorithms also through simulation.  

\end{abstract}
 
%%%%%%%%%%%%%%%%%%
\section{Introduction}
\label{Introduction}

Wireless infrastructureless networks include sensor- and actor-networks, wireless mesh networks, and agent networks. Such networks 
are composed of a physical and a logical configuration (topology). A physical configuration is characterized by node positions and connectivity. 
A logical configuration is characterized by link activities, i.e., a pattern of node-node communications on who is communicating with whom and when. For wireless networks, both a physical and a logical configuration can vary due to either failures or environmental changes. Configuration management is to adapt a physical and/or a logical topology to support and maintain node-node communications. 

There is no centralized management authority for infrastructureless wireless networks. Self-configuration is desirable for nodes to adjust adaptively their own positions and communication \textcolor{black}{activities} in a distributed setting through local interactions. A challenge is whether distributed self-configuration would result in a near-optimal configuration with a sufficiently small approximation error. 

In this work, we develop an analytical model for distributed self-configuration, and study the issue of near-optimality. To be specific, we consider joint formation and re-configuration of a physical topology and scheduling upon node failures.  For simplicity, we focus on ad-hoc wireless networks \textcolor{black}{with node failures and random perturbations of positions but} without mobility. \textcolor{black}{Our prior work \cite{Mass05} developed an initial probabilistic graphical model for the distributed configuration.  This work enhances modeling, analysis on the performance-complexity trade-off, and validation through simulation.}  

What is optimality? Deterministic optimization has been used to obtain an optimal solution for centralized management \cite{Bjorklund}\cite{Chiang}\cite{Elbatt}\cite{Madan}. A cost function is derived to consist of management objectives and constraints. Such an approach has been applied to network capacity maximization through adapting a physical topology \cite{Greiner}, and to link-scheduling through configuring a logical topology \cite{Bjorklund}. \textcolor{black}{Open issues are how to include random factors such as inaccurate node positions, wireless channels, interference, and locally interacting wireless nodes. Another open issue is whether a cost function itself is optimal, which may go beyond a conventional optimization framework.} 

From a computational standpoint, global optimization requires a centralized entity to maintain and update complete information for all nodes in the network. This is impractical for large networks. More importantly, locality can be \textcolor{black}{a generic feature} to configuration management. For example, in a large wireless network, nodes and links often fail locally. A local repair is thus desirable for preventing an entire network from an incessant re-configuration. Therefore, distributed configuration management is a necessity for large wireless networks \cite{Bjorklund}\cite{Mass05}.  

In a distributed setting, each node either adjusts its own physical position or decides when and whom to transmit based on local information. This should be done in a fully asynchronous and distributed fashion with local information exchange. Local information on a configuration may include physical locations and communication activities (channel-access) of neighbors. Such information can be either sensed locally at a node or exchanged with neighbors. The range of information-exchange characterizes communication complexity. When the information-exchange is performed only among close neighbors, the resulting distributed algorithm would scale gracefully with a network size. 

Numerous distributed algorithms and protocols have been developed for topology formation \cite{Spears}\cite{Steenstrup}
using local information (see \cite{Wattenhofer} and references therein). Self-organizing protocols have been developed for sensor networks \cite{Estrin}\cite{Pottie} and p2p self-stabilizing networks (see \cite{Ko} and references therein). These distributed algorithms are deterministic. \textcolor{black}{\cite{Oikonomou} uses a simple randomized algorithm to analyze throughput for different traffic conditions of wireless ad-hoc networks. \cite{Srinivasan} develops a fast randomized and distributed algorithm for edge coloring, assuming a bounded nodal degree on bipartite graphs and thus Markovian dependence.}  \cite{Baras} develops a probabilistic model with spatial Markovian assumptions to characterize the randomness in node positions. 
\textcolor{black}{\cite{Modiano} uses gossiping algorithms to show that the throughput of wireless networks can be maximized in a fully distributed fashion. These randomized algorithms either assume Markovian spatial dependence which results in the optimality or require non-local information for local-decisions. \cite{Elbatt} \cite{Madan} show that more general SINR-based interference models result in non-Markovian  (long-range) spatial dependence, and  the aggregated interference from small interfering links of different strengths may not be negligible \cite{Elbatt}. \cite{Kauffman} proposes a measurement-based self-organization for 802.11 wireless access network. \cite{Kauffman} is based on a Gibbs distribution, thus the global optimality can be achieved asymptotically as shown in \cite{Geman}. However, different from 802.11 wireless access networks, nodes in ad hoc wireless networks may not have complete information especially from far-away nodes.
}

\textcolor{black}{
A question arises \cite{Modiano}, if the spatial dependence is non-Markovian, when is distributed configuration management near optimal? Thus, quantitative conditions need to be derived on {\it when} and {\it how} the distributed management can result in a near-optimal configuration.} In fact, it has been considered as a difficult problem in general to develop a distributed algorithm with a predictable performance \cite{TSI}. The open questions are

 $\circ$ {\it What} measures the optimality of centralized-configuration management, and the near-optimality of distributed-configuration management? 

 $\circ$ {\it When} is it possible for distributed management to achieve a near-optimal configuration? 

 $\circ$ {\it How} to derive distributed algorithms that obtain a near-optimal configuration?

This work intends to develop a framework to study these issues through {\it modeling}, {\it analysis}, and {\it algorithms}.\\

{\bf Global Model and Optimality}: The optimality of configuration management can be considered in a model-based framework. There, if network configurations can be modeled based on the ground truth, the optimality can be defined accordingly. Centralized configuration management can be used to define the optimality of a network configuration as nodes are not limited by the scope of information exchange. 

We consider three factors in a model of configuration: Randomness resulting from a network internally, management constraints imposed externally, and distributed decisions made by nodes with limited information. Randomness in wireless networks is challenging to model \cite{Mass05}. This work begins with simple scenarios where the  randomness results from perturbed node positions, interference, and node-node communication. Fading is not considered in this work for simplicity. Management constraints include requirements on the connectivity of a physical topology, signal-to-interference plus noise ratio (SINR) and reconfiguration cost. Nodes make asynchronous and randomized decisions,  adjusting a configuration in a distributed setting. The model is developed from bottom-up by mapping these three factors onto a probability distribution. The model is thus accurate in regard to the ``ground truth" based on the assumed wireless channel, constraints and nodal decisions. Overall, such a model characterizes statistical spatial dependence of nodes/links in a network configuration. 

To obtain an analytical form of the probabilistic model, we adopt an analogy between link activities and node positions of a wireless network and interacting particles in statistical physics \cite{Huang}\cite{Wagner}. Such an analogy allows the use of ``configuration Hamiltonian" \cite{MRF} for quantifying a network configuration. The configuration Hamiltonian corresponds to an artificial system energy of a wireless network. The system energy combines the physical topology, link activities, and management constraints into a single quantity. The configuration Hamiltonian is then used to obtain a probabilistic model which is known as a Gibbs distribution \cite{Huang}. Such a Gibbs distribution is for an entire ``network" configuration and thus corresponds to a global probabilistic model.\\ 

\indent
 {\bf Local Model and Probabilistic Graphs:} We obtain a local model as an approximation of the global model of network configurations. 
To approximate the global model,   we rely on a one-to-one mapping between a Gibbs distribution and a probabilistic graphical model in machine learning \cite{Geman}\cite{Jordan}.  The graph provides a simple and explicit representation of statistical spatial dependence in a network configuration. 

We show that a probabilistic graph of the global model belongs to a two-layer random-field. One layer is for a physical configuration, and the other layer is for a logical configuration. The graph is fully connected \textcolor{black}{, i.e., non-Markovian}, where the long-range spatial dependence results from the interference among far-away nodes. When the long-range interference can be neglected, the global model can be approximated by a two-layer coupled Markov Random Field which is also called a Random Bond model. The corresponding dependency graph exhibits a nested spatial Markov dependence for both a physical and  logical configuration. Mathematically, such a spatial Markov dependence can be represented as a product of local conditional probability density functions \cite{Geman}. Hence the probabilistic graphical model shows which ``dependency links" to remove, result in a local model.

\textcolor{black}{
We define an approximation error to measure the difference between the local and global model. When the approximation error is within a given bound, a local model is near-optimal. We then define complexity to characterize the size of a neighborhood where nodes exchange information locally.  We obtain bounds for the approximation error. The approximation error and the bounds depend on the following parameters from the physical and logical configuration: the power decay of a wireless channel, the density of active (transmitting) nodes, the complexity, and the size of a network. We quantify the impact of these interacting parameters on the approximation error for a large network.  We derive sufficient conditions for the local model to be near-optimal, i.e., when the total interference from far-away nodes decreases faster than the growth of the number of interfering nodes as the network size increases.
}\\ 

\indent
{\bf Distributed Algorithm}: A local model results in a class of distributed algorithms where nodes self-configure through information exchange with neighbors. The range of information exchange characterizes the local connectivity of the probabilistic dependency graph, and corresponds to the communication complexity. The actual information exchanged includes relative positions of neighbors and activities of adjacent nodes. Node decisions are probabilistic, corresponding to randomized distributed algorithms of graphical models.\\
 
\indent 
We apply the distributed algorithm to three examples of self-configuration: (a) forming a 1-connected physical topology from a random initial topology; (b) configuring a logical topology that maximizes spatial channel-reuse and incessant communication demands; and (c) reconfiguring jointly a physical- and logical-configuration upon failures. \textcolor{black}{We use simulation to validate the models and bounds, and demonstrate how the framework enables fully distributed spatial scheduling algorithms and fault tolerance for wireless infrastructureless networks.}

%%%%%%%%%%%%%%%%%%
\section{Problem Formulation}
\label{PF}

\subsection{Assumptions}
\label{Assumptions}

Consider a wireless network with the following assumptions.

{\it Physical Layer:} All nodes share a common frequency channel. A pair of nodes within a communication range can communicate directly with an omni-directional antenna. The wireless channel follows a path-loss model with a power attenuation of factor $\alpha$ $\in$ \{$2 \sim 6$\}. Shadowing and/or multi-path fading are not considered in this work for simplicity. Node $i$  transmits with power $P_i$, where $0 \le P_i \le P_{max}$, $1 \le i \le N$, with $P_{max}$ being the maximum transmission power, and $N$ being the number of nodes in the network. Power control is not considered in this work.

{\it MAC Layer:} Let $\mbox{SINR}_{th}$ be a given threshold for the SINR requirement.  Node $i$ can transmit to node $j$ when the SINR requirement is satisfied, i.e., $\mbox{SINR}_{ij}$ = ${P_i l_{ij}^{-\alpha} \over N_b + \sum_{(m,n) \ne (i,j) } P_m l_{mj}^{-\alpha}}$ $\ge$ $\mbox{SINR}_{th}$, where $l_{ij}$ is the distance between nodes $i$ and $j$, and $N_b$ is the noise power.  We consider a scheduled resource allocation that is implemented with local interactions among neighbors.

{\it Configuration Management:} Let $X_{i0}$ and $X_i$ be a desired and an actual location of node $i$, for $1 \le i \le N$. $\mbox{\boldmath$X$} = \{ X_1, \cdots, X_N \}$ are random positions of nodes in a network where the randomness results from perturbed positions, random movements and measurement errors. 

Let $\sigma_{ij}$ denote channel-access of link ($i,j$), where $\sigma_{ij} =1$ if node $i$ is transmitting to node $j$; and $\sigma_{ij}= -1$, otherwise. $\sigma_{ij}$ is referred to as a ``communication dipole" in this work, for $1 \leq i,j \leq N$, $i \neq j$, and $\mbox{\boldmath$\sigma$}$ = $\{ \sigma_{{}_{1,2}}, \cdots, \sigma_{{}_{N,N-1}} \}$ denotes a set of link activities in the network. Link activities are assumed to be random as they are triggered by network-layer random traffic demands. A logical configuration  is $\mbox{\boldmath$\sigma$}$ = $\{\sigma_{ij}\}$. A network configuration is ($\mbox{\boldmath$\sigma$}, \mbox{\boldmath$X$}$).\\

The objectives are to achieve distributed configuration management, i.e., to (a)  form a desired physical topology, (b) schedule the resource utilization at a given time to maximize the spatial channel-reuse with a desired SINR requirement, and (c) reconfigure upon failures by minimizing reconfiguration cost.

%%%%%%%%%%%%%%%%%%%%%%
\subsection{Formulation}
\label{Formulation1}

Let P($\mbox{\boldmath$\sigma$},\mbox{\boldmath$X$}$) be a true probabilistic global model of a network configuration that results from the above assumptions.\\

\indent
Definition 1: {\it Optimal Configuration. ($\mbox{\boldmath$\sigma$}^*, \mbox{\boldmath$X$}^*$) is an optimal configuration if it maximizes the global likelihood, 
\begin{equation}
\label{GOpt}
(\mbox{\boldmath$\sigma$}^*,\mbox{\boldmath$X$}^*) =  
\mbox{arg} \mathop{\mbox{max}}_{(\mbox{\boldmath$\sigma$},\mbox{\boldmath$X$})}   P(\mbox{\boldmath$\sigma$},\mbox{\boldmath$X$}).
\end{equation}
}

Let $P^l(\mbox{\boldmath$\sigma$},\mbox{\boldmath$X$})$ be a local model that approximates the global model P($\mbox{\boldmath$\sigma$}$).\\   

\indent
Definition 2: {\it Near-Optimal Configuration. Consider  ($\hat{\mbox{\boldmath$\sigma$}},\hat{\mbox{\boldmath$X$}}$) that maximizes $P^l(\mbox{\boldmath$\sigma$},\mbox{\boldmath$X$})$, i.e., 
\begin{equation}
(\hat{\mbox{\boldmath$\sigma$}},\hat{\mbox{\boldmath$X$}}) = \mbox{arg} \mathop{\mbox{max}}_{(\mbox{\boldmath$\sigma$},\mbox{\boldmath$X$})}   P^l(\mbox{\boldmath$\sigma$},\mbox{\boldmath$X$}). 
\end{equation}
Consider an approximation error as the average relative difference between the log likelihoods, 
\begin{equation}
\label{E_Delta_error}
E[\Delta] = E \left [ \left |{{\log P(\mbox{\boldmath$\sigma$}^*,\mbox{\boldmath$X$}^*)-\log P(\hat{\mbox{\boldmath$\sigma$}},\hat{\mbox{\boldmath$X$}})} \over {\log P(\mbox{\boldmath$\sigma$}^*,\mbox{\boldmath$X$}^*)}} \right | \right ], 
\end{equation}
where the expectation is over $\hat{X}$, $X^*$, $\sigma^*$, and $\hat{\sigma}$. 
For a given $\epsilon >0$, if $E[\Delta] \le \epsilon$, 
\textcolor{black}{
the local model $P^l(\mbox{\boldmath$\sigma$},\mbox{\boldmath$X$})$ and the corresponding realization  ($\hat{\mbox{\boldmath$\sigma$}},\hat{\mbox{\boldmath$X$}}$) are 
near-optimal.}
}

%CHECK: there is a problem with defining 
%($\hat{\mbox{\boldmath$\sigma$}},\hat{\mbox{\boldmath$X$}}$) to be near-
%optimal as the E[] averages over 
%($\hat{\mbox{\boldmath$\sigma$}},\hat{\mbox{\boldmath$X$}}$) already, 
%and is no long a function of 
%($\hat{\mbox{\boldmath$\sigma$}},\hat{\mbox{\boldmath$X$}}$). 

Distributed configuration management requires that $P^l(\mbox{\boldmath$\sigma$},\mbox{\boldmath$X$})$ is factorizable, i.e., $P^l(\mbox{\boldmath$\sigma$},\mbox{\boldmath$X$})$=$\prod_{ij}$ $g_{ij}$($\mbox{\boldmath$\sigma$},\mbox{\boldmath$X$}$), where $g_{ij}$($\mbox{\boldmath$\sigma$},\mbox{\boldmath$X$}$) is a localized probability density function that depends on variables in a neighborhood of nodes $i$, $j$ and link ($i,j$) for $1 \le i,j \le N$. The global maximization from Eq.(\ref{GOpt}) reduces to a set of coupled local maximizations, i.e., 
($\hat{\sigma}_{ij},\hat{X}_i$) = $\mbox{arg}$ $\mbox{max}_{(\sigma_{ij},X_i)}$   $g_{ij}(\mbox{\boldmath$\sigma$},\mbox{\boldmath$X$})$ for $1 \le i,j \le N$.

%Sung-eok, I've removed a few large spacings like this.  

\textcolor{black}{
Note that when variables exhibit spatial Markovian dependence, distributed configuration management is naturally optimal\cite{Geman}. But when variables are non-Markovian, the near-optimality becomes important. Hence, $E[\Delta]$ provides a performance bound of distributed configuration management. If $E[\Delta]$ is less than a given error, the near-optimality can be achieved by a fully distributed  configuration management based on co-operations of nodes in a small neighborhood.\\
}

\indent
Our tasks are to 

(a) obtain a global model P($\mbox{\boldmath$\sigma$},\mbox{\boldmath$X$}$) from the above given assumptions; 

(b) characterize the spatial dependence of multiple variables ($\mbox{\boldmath$\sigma$}, \mbox{\boldmath$X$}$) using a graphical representation of   P($\mbox{\boldmath$\sigma$},\mbox{\boldmath$X$}$). Obtain a simplified graph and a mathematical representation for $P^l$($\mbox{\boldmath$\sigma$}$,$\mbox{\boldmath$X$}$);   

(c) obtain sufficient conditions for $P^l($\mbox{\boldmath$\sigma$},\mbox{\boldmath$X$}$)$ to result in a near-optimal configuration;  

(d) derive a distribution algorithm, and  apply the algorithm to self-configuration.\\

\textcolor{black}{
\indent
Table \ref{Notations_table} summarizes key notations used in the paper. 
}

\textcolor{black}{
{\small
\begin{table}[htb!]
\caption{Important Notations}
{\small
\begin{center}
\begin{tabular}{l|l}
\hline \hline
$\alpha$ & Power attenuation factor, $2 \le \alpha \le 6$ \\ \hline 
$N_b$ & Power of channel noise \\ \hline
$\mbox{SINR}_{ij}$ & ${P_i l_{ij}^{-\alpha} \over N_b + \sum_{(m,n) \ne (i,j) } P_m l_{mj}^{-\alpha}}$\\ \hline
$\sigma_{ij}$ & Activity of link ($i,j$),  $\sigma_{ij}$ $\in$ $\{-1,1\}$, $\eta_{ij}$=${\sigma_{ij}+1 \over 2}$ \\ \hline
$X_i$ & Position of node $i$ \\ \hline
$E(\Delta)$ & Approximation error of the local model \\ \hline
$\epsilon_{\Delta}$ & An upper-bound of approximation error $E(\Delta)$ \\ \hline
$\epsilon$ & Desired threshold of approximation error $E(\Delta)$ \\ \hline
$Z$ & Normalization constant \\ \hline
\end{tabular}
\end{center}
}
\label{Notations_table}
\end{table}
}
}

%%%%%%%%%%%%%%%%%%
\section{Global Model}
\label{WNModel}

We begin by developing a global model that characterizes probabilistic spatial dependence in a network configuration. Our approach is bottom-up so that the probabilistic model can be obtained faithfully based on the given assumptions and management constraints. 

\subsection{Logical Configuration}

We begin with modeling a logical configuration given node positions. \textcolor{black}{A logical configuration is considered as random,  detailed explanations and an example can be found in \cite{Thesis-Jeon}.
}

%%%%%%%%%%%%%%%%%%
\subsubsection{Configuration Hamiltonian}
\label{Hamiltonian_Sigma}

We now develop a probabilistic model for  logical configuration $\mbox{\boldmath$\sigma$}$ given a set of node positions $\mbox{\boldmath$X$}$. We regard $\mbox{\boldmath$\sigma$}$ as a set of communication dipoles. Terminology ``dipole" is originally used in a particle system in statistical physics. There,  a dipole corresponds to a particle with binary states, active or inactive \cite{MRF}.  Now consider each ``communication dipole" as a particle. Table $I$ compares a wireless network with a particle system through their similarities.

\begin{table}[htb!]
\caption{Correspondence between Wireless (Dipole) Network and Particle Systems}
{\small
\begin{center}
\begin{tabular}{l|l}
\hline \hline
Wireless (Dipole) Network & Particle Systems (Lattice Gas \cite{MRF}) \\ 
\hline \hline                                                                                                                                                                                                                                                                                                                                                                                                                                                                                                                              
active(+1) / inactive(-1) & occupied(+1) / empty(-1) \\ 
\hline
interference & interaction energy \\
\hline
system potential energy & chemical potential \\
\hline
logical configuration & system state (e.g., liquid or gas) \\
\hline
\end{tabular}
\end{center}
}
\label{D_L}
\end{table}

\indent
Configuration Hamiltonian has been applied to a particle system to describe the states of a set of particles under the following conditions: (a) active particles are statistically distinguishable, and (b) interactions between particles are weak. 

We now extend the notion of configuration Hamiltonian to the wireless network, where active communication dipoles are statistically distinguishable; and interactions among dipoles are weak due to decaying interference. We define a system energy of a logical configuration as the summation of the received power at individual receivers in the network,   
\begin{equation}
\sum_{ij} P_j \cdot {\sigma_{ij} + 1 \over 2},
\end{equation}
\noindent
where $P_j$ denotes the net received-power at the receiver $j$ by considering the signal component, interferences and noise. 
\textcolor{black}{
Note that for a link ($i$,$j$) where node $i$ is a transmitter and $j$ is a receiver, $P_i$ denotes the transmission power of transmitter $i$, and $P_j$ denotes the received power at receiver $j$. We define the received power $P_j$ to include the transmitted signal from the transmitter $i$ and the interference from interferers with the opposite sign of the signal from transmitter $i$.
}
Based on the assumptions in Section \ref{PF}, for a single active dipole $\sigma_{ij}=1$ in the network, the received power at the receiver $P_j$ = $P_i l_{ij}^{- \alpha}$ ${ \sigma_{ij} + 1 \over 2}$, where $l_{ij}=|X_i-X_j|$. A dipole is inactive, i.e., $\sigma_{ij}=-1$ and $P_j=0$, if node $i$ does not transmit to node $j$. For multiple active dipoles, $P_j$ has addition terms as the interference. 
%Sung-eok: write out the interference terms also.

Following the definitions in statistical physics \cite{Huang},
the ``configuration Hamiltonian" of a dipole system is the negative system energy \cite{Mass05}, 
\begin{equation}
\label{Power4_1}
H(\mbox{\boldmath$\sigma$}|\mbox{\boldmath$X$}) =
   -\sum_{ij} P_j \eta_{ij} + \beta \sum_{ij} (\mbox{SINR}_{ij} - \mbox{SINR}_{th})^2  \eta_{ij},
\end{equation}
\noindent
where $\eta_{ij}$ = ${\sigma_{ij} + 1 \over 2}$,
{\small $\mbox{SINR}_{ij}$ = ${P_i l_{ij}^{-\alpha} \eta_{ij} \over \sum_{mn \neq ij} P_m l_{mj}^{-\alpha} \eta_{mn} + N_b}$} is the SINR for dipole $\sigma_{ij}$, $\mbox{SINR}_{th}$ is a given SINR threshold, and $\beta>0$ is a weighting factor. %
$\beta (\mbox{SINR}_{ij}$ $-$ $\mbox{SINR}_{th})^2$ serves as a penalty term for the SINR constraint\footnote{The use of a quadratic function as the constraint is for simplicity of derivation. A hard constraint is used in the distributed algorithm, i.e., $\beta U(\mbox{SINR}_{th}$ $-$ $\mbox{SINR}_{ij})$ where $U(x)=1$ for $x>0$; 0, otherwise.}, an equivalence of which is used for simplicity, i.e., {\small $\beta [ P_i l_{ij}^{-\alpha} \eta_{ij} - \mbox{SINR}_{th} ( \sum_{mn \neq ij} P_m l_{mj}^{-\alpha} \eta_{mn} + N_{b_{ij}})]^2$}.\\

\indent
For an active dipole $\sigma_{ij}=1$, the interference sources within a certain neighborhood from the receiver $j$ are considered as the significant interferers; and this neighborhood is denoted by $N_{ij}^I$ as the interference range of node $j$. The Hamiltonian can be rewritten as 
{\small
\begin{equation}
\label{H1}
H(\mbox{\boldmath$\sigma$}|\mbox{\boldmath$X$}) = R_1(\mbox{\boldmath$\sigma$},\mbox{\boldmath$X$}) 
+ R_2(\mbox{\boldmath$\sigma$},\mbox{\boldmath$X$})                                         + R_3(\mbox{\boldmath$\sigma$},\mbox{\boldmath$X$}) 
+ R_I(\mbox{\boldmath$\sigma$},\mbox{\boldmath$X$}), 
\end{equation}
}
\noindent
where 
{\small $R_1(\mbox{\boldmath$\sigma$},\mbox{\boldmath$X$})$ = $\mathop{\sum}_{ij} \alpha_{{}_{ij}} \eta_{ij}$} is the first-order energy of individual dipoles, 
{\small $R_2(\mbox{\boldmath$\sigma$},\mbox{\boldmath$X$})$ = $\mathop{\sum}_{ij} \mathop{\sum}_{\mbox{{\tiny $mn \in N_{ij}^I$}}}$ $\alpha_{{}_{ij,mn}} \eta_{ij} \eta_{mn}$} is the second-order energy with products of two dipoles within the interference range,
{\small $R_3(\mbox{\boldmath$\sigma$},\mbox{\boldmath$X$})$ = $\mathop{\sum}_{ij} \mathop{\sum}_{\mbox{{\tiny $mn \in N_{ij}^I$}}}$ 
$\mathop{\sum}_{\mbox{{\tiny $uv \in \{N_{ij}^I,N_{mn}^I\}$}}}$ $\alpha_{{}_{ij,mn,uv}} \eta_{ij} \eta_{mn} \eta_{uv}$} is the third-order energy with products of three dipoles within the interference range,
{\small $R_I(\mbox{\boldmath$\sigma$},\mbox{\boldmath$X$})$ = $\sum_{ij} R_{I_{ij}}(\mbox{\boldmath$\sigma$},\mbox{\boldmath$X$})$} is the total interference outside the interference range where 
{\small $R_{I_{ij}}(\mbox{\boldmath$\sigma$},\mbox{\boldmath$X$})$} is the residual interference outside the interference range of an active dipole $\sigma_{ij}=1$. 
The coefficients of the link activities $\mbox{\boldmath$\sigma$}$ depend on  relative node positions $l_{ij}$'s, where
{\small
\begin{eqnarray}
\label{Alpha1}
\alpha_{ij}     &=& {-P_i l_{ij}^{-\alpha}}  + \beta \cdot (P_i l_{ij}^{-\alpha} - \mbox{SINR}_{th} N_b)^2,  \\ 
\alpha_{ij,mn}  &=& {2 \sqrt{P_i P_m} l_{ij}^{-{\alpha \over 2}} l_{mj}^{-{\alpha \over 2}}  - P_m l_{mj}^{-\alpha}}
                    +  \beta \mbox{SINR}_{th}^2 P_m^2 l_{mj}^{-2\alpha}                                                  \nonumber \\ \nonumber
                & & -2 \beta (P_i l_{ij}^{-\alpha} - \mbox{SINR}_{th} N_b) \cdot \mbox{SINR}_{th} P_m l_{mj}^{-\alpha},  \nonumber \\ \nonumber
\alpha_{ij,mn,uv} &=& {- 2 \sqrt{P_m P_u} l_{mj}^{-{\alpha \over 2}} l_{uj}^{-{\alpha \over 2}}} 
                      + \beta (\mbox{SINR}_{th}^2 P_m P_u l_{mj}^{-\alpha} l_{uj}^{-\alpha}). \nonumber
\end{eqnarray}
}
Intuitively, $\alpha_{ij}$ corresponds to the increased power when dipole $\sigma_{ij}$ becomes active, $\alpha_{ij,mn}$ relates to the interference experienced  by $\sigma_{ij}$ resulting from a neighboring active dipole $\sigma_{mn}$, and $\alpha_{ij,mn,uv}$ relates to the interference experienced by $\sigma_{ij}$ from both $\sigma_{mn}$ and $\sigma_{uv}$.

%%%%%%%%%%%%%%%%%%%%%
\subsection{Physical Configuration}
\label{Hamiltonian_X}

Now consider node positions $\mbox{\boldmath$X$}$. $\mbox{\boldmath$X$}$ is assumed to be random where the randomness originates from perturbed node locations, e.g., due to random movements from desired locations and measurement noise. \textcolor{black}{For example, a set of desired positions can be pre-determined as a management constraint to form a regular grid. But the actual node positions may deviate from their desired positions, resulting in an irregular topology.}

Management constraints are imposed on the physical connectivity. The 1-connectivity is an example where there exists at least one connected path between any two nodes in the network to achieve the reachability of any source-destination pair. A Yao-graph provides a sufficient condition of the 1-connected physical topology, where each node has a connected link with its nearest neighbors every $\theta$ ($\leq {2 \pi \over 3}$) radian apart \cite{Wattenhofer}. Such a constraint can be represented as  
\begin{equation}
\label{Yao0}
h(X_i, X_j) = \left \{ \noindent
  \begin{array}{ll}
  0, & |{l_{ij}-l_{th} \over l_{th}}|   \le  \epsilon_0, \\
  |l_{ij}-L_{ij}|, & \mbox{otherwise,}\\
  \end{array}
\right.\
\end{equation}
%check here
where $\epsilon_0$ is a small positive constant, $L_{ij}$=$|X_{i0}-X_{j0}|$ is the desired distance \footnote{When $L_{ij}$ = $l_{th}$ for a positive constant $l_{th}>0$, \textcolor{black}{resulting in equally placed nodes and a regular Yao-graph}.} of $l_{ij}$, $ X_{i0}$ is a desired position of node $i$, \textcolor{black}{which can be pre-determined as management objectives, desired positions,  or the initial positions to minimize changes  in positions},  $\forall$ ($i,j$). The resulting Hamiltonian for the physical topology is 
{\small
\begin{eqnarray}
\label{UH}
H(\mbox{\boldmath$X_{}$}) &=& \sum_i {(X_i - X_{i0})^2 \over 2 \sigma^2} + \zeta \sum_i \sum_{j \in N_i^{\theta}} h(X_i, X_j),
\end{eqnarray}
}
where $N_i^{\theta}$ is the set of the nearest neighbors of node $i$ for every $\theta$ radian, $\sigma$ is the variance which is assumed to be the same for all nodes for simplicity, and $\zeta$ is a positive weighting constant.

%%%%%%%%%%%%%%%%%%%%%%%%%
\subsection{Network Configuration}
\label{Joint}

We now consider a network configuration which consists of both a physical and a logical configuration. 
%%%%%%%%%%%%%%%%%%%%%%%%%%%%%%%%%%%%%%%%%%%%%
\subsubsection{Network Configuration Hamiltonian}

Combining the Hamiltonians from the physical and logical configurations results in an overall network-configuration Hamiltonian, 
\begin{equation}
\label{varsig}
H(\mbox{\boldmath$\sigma_{}$}, \mbox{\boldmath$X_{}$}) = 
 H(\mbox{\boldmath$\sigma_{}$} | \mbox{\boldmath$X_{}$}) +  H(\mbox{\boldmath$X_{}$}). 
\end{equation}
%

%%%%%%%%%%%%%%%%%%%%%%%%%%%%
\subsubsection{Gibbs Distribution}
\label{Boltz}

A configuration Hamiltonian  can be related to a probabilistic model through a Gibbs distribution \cite{MRF}. 
Specifically, in a particle system \cite{Huang}, 
the effective system potential energy $H(\omega)$, known as the configuration Hamiltonian,  obeys the Gibbs (or Boltzmann) distribution \cite{Huang}, $P(\omega)$ = $Z_0^{-1} \cdot$ $\exp \left ( {-H(\omega) \over T} \right )$, 
where $\omega$ is a multi-dimensional configuration-variable, $Z_0$ is a normalizing constant and $T$ is the temperature of the particle system  \cite{Geman}\cite{Huang}.\\

\indent
{\bf Model of logical configuration:}
\indent
For a logical configuration $\mbox{\boldmath$\sigma$}$ given node positions $\mbox{\boldmath$X$}$, a Gibbs distribution $P(\mbox{\boldmath$\sigma$}|\mbox{\boldmath$X$})$ can be obtained 
using configuration Hamiltonian $H(\mbox{\boldmath$\sigma$}|\mbox{\boldmath$X$})$, 

{\small
\begin{equation}
P(\mbox{\boldmath$\sigma$}|\mbox{\boldmath$X$}) =  Z_{\sigma}^{-1} \cdot \exp  \left ({-H(\mbox{\boldmath$\sigma$}|\mbox{\boldmath$X$}) \over T} \right ),
\end{equation}
}
\noindent 
where $Z_{\sigma} = \sum_{\mbox{\boldmath$\sigma$}} \exp  \left ({-H(\mbox{\boldmath$\sigma$}|\mbox{\boldmath$X$}) \over T} \right )$ is a normalizing constant and also called the partition function \cite{Geman}. $T >0$ is the temperature in statistical physics that characterizes the stability of the system \cite{Baras}\cite{Geman}. The lower the temperature, the more stable the configuration is. $T$ is used in \cite{Geman} as a computational variable to obtain the most probable configuration (see Section \ref{SelfManagement} for details).\\ 

\indent
{\bf Model of physical configuration:}
\indent
Similarly, the Gibbs distribution of node positions can be obtained as 
{\small
\begin{eqnarray}
P(\mbox{\boldmath$X_{}$})
&=& {Z_X}^{-1} \cdot \exp \left ( {{-H(\mbox{\boldmath$X_{}$}) \over T}} \right ),
\end{eqnarray}
}
\noindent
where $Z_X$=$\sum_{\mbox{\boldmath$X$}} \exp  \left ({-H(\mbox{\boldmath$X$}) \over T} \right )$ is a normalizing constant.\\

%*Is this T the same as the previous one?

\indent
{\bf Two-layer network model:}
\indent
The Gibbs distribution of an entire network configuration can be obtained using the overall configuration Hamiltonian, which is 
{\small
\begin{equation}
P(\mbox{\boldmath$\sigma_{}$}, \mbox{\boldmath$X_{}$})
= {Z_0}^{-1} \cdot \exp \left ({-H(\mbox{\boldmath$\sigma_{}$}, \mbox{\boldmath$X_{}$}) \over T} \right ),
\end{equation}
}
\noindent
where $Z_0$=$\sum_{(\mbox{\boldmath$\sigma$},\mbox{\boldmath$X$})} \exp  \left ({-H(\mbox{\boldmath$\sigma$},\mbox{\boldmath$X$}) \over T} \right )$  is a normalizing constant.\\
%

%%%%%%%%%%%%%%%%%%%%%%%%%%%%%%%%%%%%%%%%%%%%%%%
\subsubsection{Minimum Hamiltonian and Optimal Configuration}
\label{Opt}

An optimal configuration maximizes the likelihood function, 
\textcolor{black}{
\begin{eqnarray}
({\mbox{\boldmath$\sigma$}}^*, {\mbox{\boldmath$X$}}^*) 
&=& \mbox{arg} \mathop{\mbox{max}}_{\tiny (\mbox{\boldmath$\sigma$}, \mbox{\boldmath$X$})} P(\mbox{\boldmath$\sigma_{}$},\mbox{\boldmath$X_{}$})  \\ \nonumber
&=& \mbox{arg} \mathop{\mbox{min}}_{\tiny (\mbox{\boldmath$\sigma$},\mbox{\boldmath$X$})} H(\mbox{\boldmath$\sigma_{}$},\mbox{\boldmath$X_{}$}), 
\end{eqnarray}
}
\noindent
where $H(\mbox{\boldmath$\sigma_{}$},\mbox{\boldmath$X_{}$})$ = $-\log \left [P(\mbox{\boldmath$\sigma_{}$},\mbox{\boldmath$X_{}$})\right ]/T$ $-\log \left ( Z_0 \right )$. Note the system energy $H(\mbox{\boldmath$\sigma_{}$},\mbox{\boldmath$X_{}$})$ incorporates both the randomness and the external management requirements. When the constraints are satisfied, the penalty terms should be diminishing. Therefore, an optimal configuration should satisfy the management objectives of spatial reuse and the constraints.

%%%%%%%%%%%%%%%%%%
\section{Local Model}
\label{Graph3}

We now seek a local model $P^l(\mbox{\boldmath$\sigma$}, \mbox{\boldmath$X$})$ that is factorizable as a product of localized probability density functions, and \textcolor{black}{also} a good approximation to the global model $P(\mbox{\boldmath$\sigma$},\mbox{\boldmath$X$})$. We resort to probabilistic graphical models.

%%%%%%%%%%%%%%%%%%
\subsection{Graphical Representation}
\label{Sufficient}

Probabilistic graphical models relate a probability distribution with a dependency graph of the corresponding random variables \cite{Geman}\cite{Jordan}\cite{Kschischang}. A node in the graph represents a random variable and a link between two nodes characterizes their statistical dependence. In particular, a set of random variables $\mbox{\boldmath$v$}$ forms Gibbs Random Field (GRF) if it obeys a Gibbs distribution \cite{MRF}. Hammersley-Clifford theorem shows an equivalence between a probabilistic dependency graph and a Gibbs distribution.\\

\indent
Hammersley-Clifford Theorem \cite{MRF}: {\it Let $S = \{1, \cdots, N\}$ be a set of nodes, and $\mbox{\boldmath$v$}$ be a set of random variables $\mbox{\boldmath$v$} = \{ v_1, \cdots, v_N \}$. $\mbox{\boldmath$v$}$ is said to be a Markov Random Field if (i) P($\mbox{\boldmath$v$}$) $>$ 0 for $\forall$ $\mbox{\boldmath$v$}$ in sample space; (ii) $P(v_i| \mbox{\boldmath$v_j$} \; \mbox{for} \; j \in S \backslash \{i\})$ = $P(v_i| \mbox{\boldmath$v_j$} \; \mbox{for} \; j \in N_i)$, where $N_i$ is a set of neighbors of node $i$ for $1 \le i \le N$.\\
\indent
Random field $\mbox{\boldmath$v$}$ is also a Gibbs Random Field if its probability distribution can be written in a product form P($\mbox{\boldmath$v$}$) = $\prod_{c \in C} V_c(\mbox{\boldmath$v$})$, where $c$ is a clique, $\mathcal{C}$ is the set of all feasible cliques, and $V_c(\mbox{\boldmath$v$})$ is a positive function.\\
}

\begin{figure}[htb] \centering
  \begin{tabular}{c}
    \resizebox{9.0cm}{!}{\includegraphics{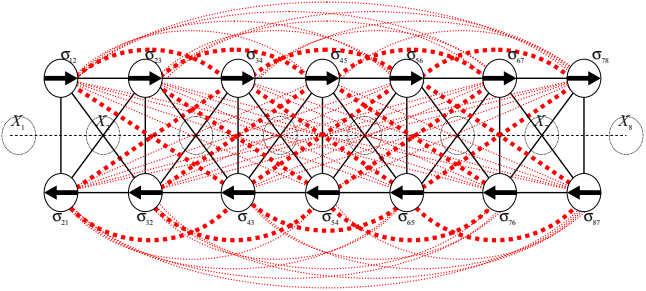}} % \\ %M2.eps
   \end{tabular}
\caption{Dependency Graph of Link Activities $\mbox{\boldmath$\sigma$}$ given Node Positions $\mbox{\boldmath$X$}$ for a line network. Contention range is given to be one hop and interference range is two hops. }
\label{DSigma}
\end{figure}

\indent
Hammersley-Clifford Theorem asserts that $\mbox{\boldmath$v$}$ is a Markov Random Field {\it if and only if} the probability distribution P($\mbox{\boldmath$v$}$) follows a Gibbs distribution with conditional independence given neighbors. This theorem shows an interesting type of probabilistic graphical models where a random variable is conditionally independent of the others given its neighbors. In particular, when the neighborhood is much smaller than the size of a network, the conditional independence implies an interesting type of spatial Markov dependence, i.e., a node depends on its far neighbors through neighbors' neighbors. Such a nested dependence can be shown explicitly through local connectivities among nodes in a dependency graph. The resulting probability distribution is thus factorizable in terms of local probability distributions.\\

%Sung-eok: move Fig. 1 to here or at the end of this sub-section.
%Also, may truncate parts of Fig. 1 to make it more clear. (Ok if it
%Only shows a part the real thing.)
%
\indent
For example, consider a one-dimensional physical topology, and the corresponding dependency graph for link activities $\mbox{\boldmath$\sigma$}$ given $\mbox{\boldmath$X$}$ shown in Figure \ref{DSigma}. Nodes in the graph represent binary random variables $\sigma_{ij}$'s, and the links represent their spatial dependence. For example, solid lines show the dependence due to channel contention, the thin dashed lines indicate the dependence due to exact interference, and the thick dashed lines correspond to an approximation of the exact spatial dependence. The first and second rows of dipoles show the activity of bidirectional links.  

All communication dipoles are fully connected due to interference, and thus exhibit non-Markovian spatial dependence. Such spatial dependence is represented by the Hamiltonian in Eq.(\ref{H1}) that consists of the products of all dipole-pairs. The fully connected dependency graph shows an uninteresting case of a random field where the neighborhood of a node is the entire network. This implies that obtaining an optimal configuration would require each node to exchange information with all the other nodes in the network.

%%%%%%%%%%%%%%%%%%%%%
\subsection{Approximation}

\textcolor{black}{To our knowledge, little has been done in the prior work on how to obtain an approximation of a fully connected dependency graph for near-optimal distributed configuration management.} We observe that the interference outside the interference range, $R_I(\mbox{\boldmath$\sigma$},\mbox{\boldmath$X$})$,
can be relatively small compared to the first three terms of the configuration Hamiltonian in Eq.(\ref{H1}).
The third-order term, $R_3(\mbox{\boldmath$\sigma$},\mbox{\boldmath$X$})$,
can be small also compared to the second-order term.
\textcolor{black}{That is, in Eq.(\ref{H1}), for an active link ($i,j$), the third terms include $l_{ij}^{-\alpha/2}  l_{mj}^{-\alpha/2} l_{uj}^{-\alpha/2}$ where $l_{mj} >> l_{ij}$ and $l_{uj} >> l_{ij}$; whereas, the second term is only the product of $l_{ij}^{-\alpha/2} l_{mj}^{-\alpha/2}$. } 
Hence, if we use the first two terms to approximate the configuration Hamiltonian, we have 
{\small
\begin{eqnarray}
\label{pl}
H^{l}(\mbox{\boldmath$\sigma$}|\mbox{\boldmath$X$}) =
\mathop{\sum}_{ij} \alpha_{{}_{ij}} (\mbox{\boldmath$X$}) \eta_{ij}
+ \mathop{\sum}_{ij} \mathop{\sum}_{mn \in N_{ij}^I} \alpha_{{}_{ij,mn}} (\mbox{\boldmath$X$}) \eta_{ij} \eta_{mn},  
\end{eqnarray}
}
\noindent
and the corresponding Gibbs distribution is
\begin{eqnarray}
\label{pl1}
& & P^l(\mbox{\boldmath$\sigma$}|\mbox{\boldmath$X$}) = Z_{l}^{-1} \cdot \exp \left ({- H^{l}(\mbox{\boldmath$\sigma$}|\mbox{\boldmath$X$}) \over T} \right ),
\end{eqnarray}
\noindent
where $Z_l$ is a normalization constant.\\

\noindent
As the sum in Eq.(\ref{pl}) only involves neighboring dipoles which are within the interference range, the resulting dependency graph now has a small neighborhood (see the thick dashed-lines in Figure \ref{DSigma}). In fact this approximated Markov Random Field is the well-known second-order Ising model \cite{MRF} where the Hamiltonian 
$H^{l}(\mbox{\boldmath$\sigma$}|\mbox{\boldmath$X$})$ consists of both the first- and second-order terms of $\sigma_{ij}$'s. Such an approximation can also be obtained directly from the probabilistic dependency graph of the global model. That is, by removing all edges outside the interference range of each node, we can obtain the graphical representation of the local model.

%%%%%%%%%%%%%%%%%%%%%%%%%%%%%%%%%%%%%%%%%%%%%%
\subsection{Spatial Dependence in Physical Topology}

We now examine the spatial dependence of node positions $\mbox{\boldmath$X_{}$}$. 
%**
\textcolor{black}{
In general, spatial dependence in physical topology is
not always Markovian since a general management objective can 
correspond to a fully connected graph. However, for many important and practical management objectives, a node
only needs to interact with the close neighbors. 
Thus we also consider 
a physical topology that exhibits the second-order Markov dependence  in this
work.} 

For example, under the 1-connectivity constraint shown in Eq.(\ref{Yao0}), 
node positions $\mbox{\boldmath$X_{}$}$ correspond to a second-order Markov Random Field, where the interactions are only with the first-order neighbors.

%%%%%%%%%%%%%%%%%%%%%%%%%%%%%%%%%%%%%%%%%%%%

\subsection{Two-layer Markov Random Fields}

%Sung-eok, move Fig. 2 to somewhat later in this subsection.
%Also, may truncate the fig. to fit the column. 

\begin{figure}[htb] \centering
  \begin{tabular}{c}
    \resizebox{9.0cm}{!}{\includegraphics{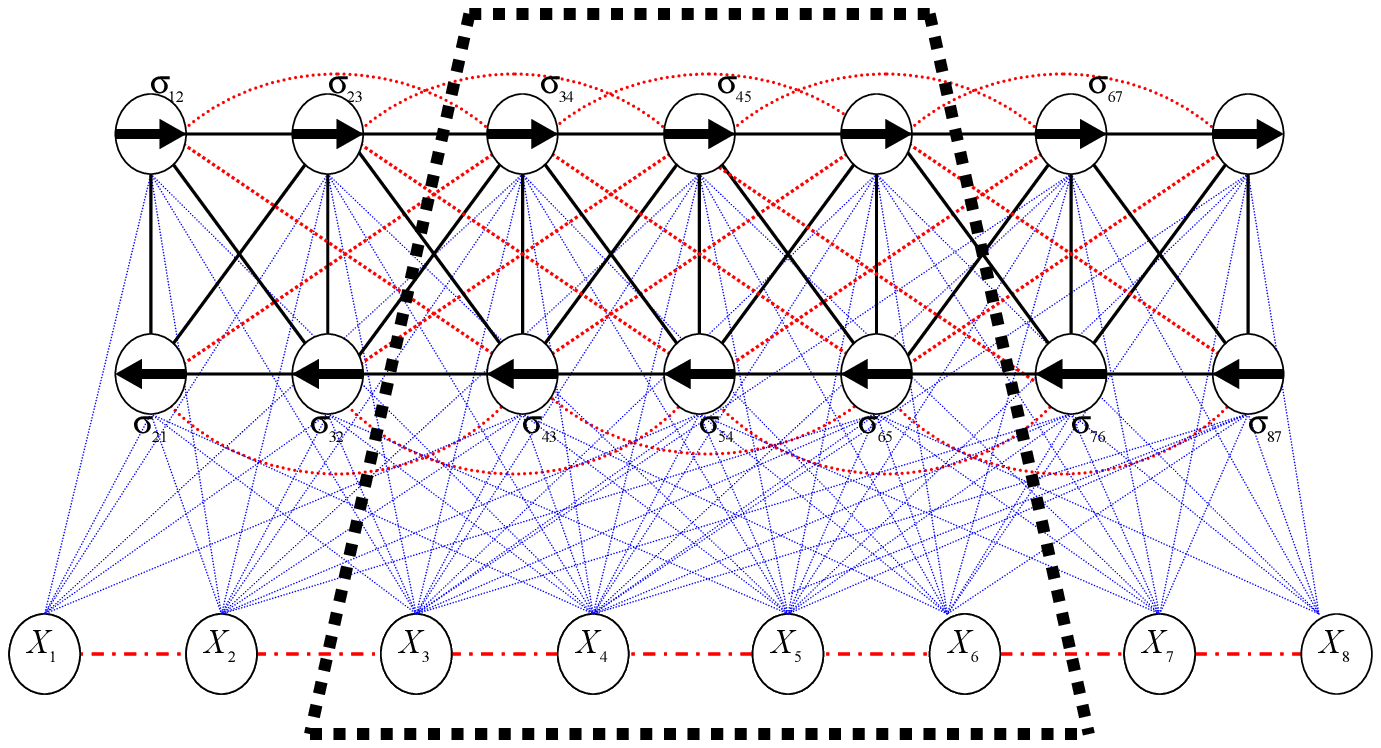}}%D3.eps \\
   \end{tabular}
\caption{Two-layer Graph and Clique of ($\mbox{\boldmath$\sigma$}$, $\mbox{\boldmath$X$}$). The Dashed-Line Box: A Clique of the Two-layer Graph}
\label{DSigma1}
\end{figure}
%
%Show clique in the figure. And resizing the fig to fit the page. 
%

The probabilistic graph of an overall configuration can be obtained by combining two graphs for the logical and physical configuration. 
For the example network, a two-layer graph is shown in Figure \ref{DSigma1} as an approximation of the original overall configuration.  The upper layer graph is for the logical configuration that assumes the local interference among neighboring active dipoles. 
The lower-layer is for the physical configuration that shows Markov dependence of node positions due to the 1-connectivity constraint in Eq.(\ref{Yao0}). The entire graph thus exhibits spatial Markov dependence at both layers. 

This two-layer graph corresponds to a coupled MRF \cite{Geman},
where an Ising model and a second-order MRF are combined together. 
The graph is also known as a Random-Bond model \cite{MRF}, where 
dipoles are connected by random bonds which depend on node positions.
The two-layer MRF ($\mbox{\boldmath$\sigma$}$, $\mbox{\boldmath$X$}$) can also be represented by a chain graph \cite{Liu05} of two MRF layers,
one for $\mbox{\boldmath$X$}$ and the other for $\mbox{\boldmath$\sigma$}$. The two-layer probabilistic graph thus maps the complex spatial dependence of a multi-hop wireless network to an explicit graphical representation. 

The corresponding likelihood function can then be represented as 
\begin{equation}
P^l(\mbox{\boldmath$\sigma$}, \mbox{\boldmath$X$}) \propto \prod\limits_{i,j} g_{ij}(\mbox{\boldmath$\sigma$}, \mbox{\boldmath$X$}),
\end{equation}
\noindent
where $g_{ij}(\mbox{\boldmath$\sigma$}, \mbox{\boldmath$X$})$ is a local probability density function and can be represented as a function of the sum of clique potentials:
\begin{eqnarray}
g_{ij}(\mbox{\boldmath$\sigma$}, \mbox{\boldmath$X$}) 
&=& \prod_{c \in C_{ij}} \exp \left ({-\psi_c( 
\mbox{\boldmath$\sigma$}, \mbox{\boldmath$X$}) \over T} \right )
    \nonumber \\ \nonumber
&=& \exp \left ({- \sum_{c \in C_{ij}} \psi_c(\mbox{\boldmath$\sigma$}, \mbox{\boldmath$X$}) \over T} \right ),
\end{eqnarray}
\noindent
where $C_{ij}$ is the set of all cliques including node $i$, node $j$, and link ($i,j$); and $\psi_c(\mbox{\boldmath$\sigma$}, \mbox{\boldmath$X$})$ is a clique potential function of a clique $c$.

As an example, consider a clique for nodes $3$ and $4$ as well as link ($3,4$) shown in Figure \ref{DSigma1}. The corresponding potential is a collection of related clique functions, i.e., 

\textcolor{black}{
$\sum_{c \in C_{34}} \psi_c(\mbox{\boldmath$\sigma$}, \mbox{\boldmath$X$})$ = 
$\left [{(X_3 - X_3(0))^2 \over 2 \sigma^2}      + 
 \zeta \sum_{j \in \{2,4\}} h(X_3,X_j) \right ]$ +
$\left [{(X_4 - X_4(0))^2 \over 2 \sigma^2}  +     
 \zeta \sum_{j \in \{3,5\}} h(X_4,X_j) \right ]$ +
$\alpha_{34}$ + 
$\sum_{mn \in \sigma_{N_{34}^I}}$ $(\alpha_{34,mn} + \alpha_{mn,34})$ 
${\sigma_{mn}+1 \over 2}$, 
}
where $\sigma_{N_{34}^I}$ = $\{ \sigma_{12},\sigma_{21}$, $\sigma_{23},\sigma_{32}$, $\sigma_{43}$, $\sigma_{45}, \sigma_{54}$, $\sigma_{56},\sigma_{65} \}$ corresponds to the set of neighboring dipoles of $\sigma_{34}$ within the interference range. The solid line connecting $\sigma_{34}$ and $\sigma_{45}$ indicates the spatial dependence due to the channel contention. The dash line connecting $\sigma_{34}$ with either $X_5$ or $X_6$ indicates the dependence of dipole $\sigma_{34}$ with positions ($X_5$,$X_6$) of a neighboring dipole $\sigma_{56}$. 
In general, a clique is determined by the interference range.%%%%%%%%%%%%%%%%%%
\section{Analysis: Near-Optimality and Complexity}
\label{LocalGlobal}

We now derive near-optimality conditions for a local model to be a good approximation of the global model. The conditions can be obtained through the approximation error (Eq. (\ref{E_Delta_error})), communication complexity, and their trade-offs.

%%%%%%%%%%%%%%%%%%%%%%%%%%%%%%%%
\subsection{Communication Complexity}

The neighborhood size in a Markov Random Field corresponds to the range of information exchange of a node with its neighbors, and thus characterizes communication complexity. Specifically, the communication complexity of an active dipole can be regarded as the maximum number of active dipoles within its interference range. As such a maximum number is random and varies from dipoles to dipoles, we use a deterministic bound for the number of active dipoles. 

\begin{figure}[htb!]
\epsfysize=2.2in
\centerline{\epsffile{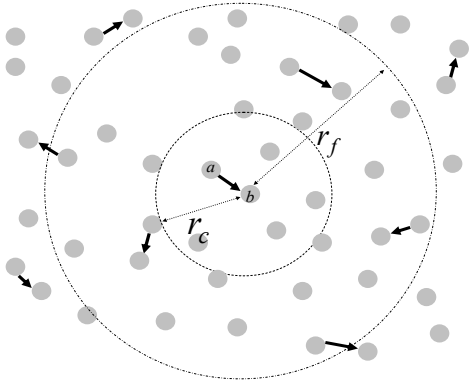}}
\caption{Contention range $r_c$ and interference range $r_f$ of an active dipole }
\label{ri_rc}
\end{figure}

Assume that an active dipole $\sigma_{ij}$ satisfies the SINR requirement, i.e., $\mbox{SINR}_{ij}$ = ${P_i l_{ij}^{-\alpha} \over N_b + \sum_{mn \neq ij} P_m l_{mj}^{-\alpha} {\sigma_{mn} + 1 \over 2}}$ $\ge$ $\mbox{SINR}_{th}$, for $1 \le, i,j, N$, $i \ne j$. 
Consider a circle centered at the receiver $X_j$ of the active dipole $\sigma_{ij}$ within which there cannot exist any active dipoles for the SINR requirement to hold. The radius of the circle is the contention range for node $j$. 
Denote the minimum contention range for all active dipoles as
$r_c$ which is the minimum distance between any two active dipoles in the network. Only one dipole can be active within a contention range. Now consider interference range outside the contention range where multiple dipoles can be active concurrently, resulting in interference. We now bound the interference region using a circular region of radius $r_f$. 
The region includes active dipoles outside $r_c$ but within $r_f$ shown in Figure \ref{ri_rc}. Note that the actual interference range of a receiver may not be symmetrical in all directions but bounded by the circular region. This circular region is now considered as the relevant interference neighborhood for node $j$. 

By packing the circular region with small circles of radius $r_c$, we can obtain the maximum number of active dipoles in the interference neighborhood.\\ 

Definition 3: {\it Communication Complexity $\mathcal{C}$.  
The communication complexity of a dipole $\sigma_{ij}$ is defined as 
the maximum number of active dipoles within the interference range, i.e., $\mathcal{C}=$ $({r_f \over r_c})^2$, for $1 \le i, j \le N$, $i \ne j$.
}

%%%%%%%%%%%%%%%%%%%%%%%%%%%%%%%%%%
\subsection{Near-Optimality Conditions}

We now derive sufficient conditions for a local model to be a good approximation of the global model. We consider homogeneous networks for ease of analysis.\\

\indent
Theorem 1: {\it 
\textcolor{black}{
Consider a network 
where $N$ nodes are uniformly distributed, and satisfies the 1-connectivity where the distance between any node and its neighbors for every $\theta$ is between 
$l_{th}(1 - \epsilon_0)$ and $l_{th}(1 + \epsilon_0)$, 
and $0 < l_{th}(1+\epsilon_0) <r_c$. Assume that $N$ nodes transmit at the same power level ($P_{t}>0$), 
have the same desired SINR threshold $\mbox{SINR}_{th}$ and the same circular interference range $r_f$. 
Let $\alpha \ge 2$ be the power attenuation factor of the channel. 
}
The average approximation error can be bounded as  E$[\Delta]$  $\leq$ $\epsilon_{{}_{\Delta}}$, where 

{\small
\begin{eqnarray}
\epsilon_{\Delta} &=
\left \{ \noindent
\begin{array}{ll}
{\mathcal{I} \over r_c^2} \left \{ (2+\ln \mathcal{C})^2 + {4\pi \over l_{th}} (  r_c+\sqrt{Nl_{th} r_c}) \right \}, & \alpha=2 \nonumber \\ \nonumber
&  \nonumber \\ \nonumber
{\mathcal{I} \over r_c^2} \left \{ {2 (2-{1 \over \mathcal{C}})^2 \over r_c^2}+{4\pi \over l_{th}^2}({1 \over \sqrt{\mathcal{C}}}+\ln(1+\sqrt{{Nl_{th} \over \mathcal{C}r_c}})) \right \}, & \alpha=4 \nonumber \\ \nonumber
& \nonumber \\ \nonumber
{\mathcal{I} \over r_c^2} \left \{ {2(\alpha-2\mathcal{C}^{2-\alpha \over 2})^2  \over (\alpha-2)^2 r_c^{\alpha-2} }  + {4\pi \over l_{th}^{\alpha \over 2}} \left [\mathcal{C}^{2-\alpha \over 4} r_c^{4-\alpha \over 2} +{2 \over 4-\alpha}  \right. \right .& \nonumber \\ \nonumber
\left . \left. (-(\sqrt{\mathcal{C}}r_c)^{4-\alpha \over 2} +(\sqrt{\mathcal{C}}r_c+\sqrt{Nl_{th}r_c})^{4-\alpha \over 2} ) \right ] \right \}, & \mbox{else}; \nonumber 
\end{array}
\right.\ \nonumber 
\end{eqnarray}
}
\noindent
with $\mathcal{I}$ = $2 l_{th}^{\alpha \over 2}/ \left(l_{th}^{-\alpha \over 2}-\sqrt{ l_{th}^{-\alpha}/\mbox{SINR}_{th}-N_b / P_{t}} \right)$.
}

\indent
The proof is given in Appendix \ref{Theorem1}. 
\textcolor{black}{The above analysis shows that a local model results in two components of the approximation error. ``Type-1 error"\footnote{We use the Type-1 and Type 2 errors for convenience and they are different from those in hypothesis testing.} is due to using a model of a lower order within interference regions, and bounded by the first terms in the above expressions. For a given communication complexity $\mathcal{C}$, the Type-1 error can be made small   by choosing a sufficiently large $r_c$. This would reduce the number of active dipoles as well as the dependence among the active dipoles. ``Type-2 error" is due to neglecting the aggregated interference outside the interference range, and upper bounded by the remaining terms in the above expressions. Type-2 error can be made small if the aggregated interference outside the interference range is reduced. Specifically, as shown by the theorem, for $\alpha > 2$ and a given $r_c$, communication complexity $\mathcal{C}$ needs to grow with respect to network size $N$ for the error to be arbitrarily small. The growth rate depends on the power decay of the channel, since the aggregated interference outside the interference range depends on $\alpha$. For a slow power decay, i.e., $\alpha =2$, the above result suggests that the Type-2 error is independent of $\mathcal{C}$, and thus can only be reduced by a sufficiently large $r_c$.}\\

\indent
{\bf Channel, contention and network size}: Using the upper bound of the approximation error $\epsilon_{{}_{\Delta}}$, we obtain a sufficient condition on the density of active dipoles for a large network so that the local model is near-optimal.\\

\indent
\textcolor{black}{
Corollary 1: 
{\it Let $\epsilon$ be a desired performance-bound. Assume $N \gg 1$, and $\mathcal{C}$ does not grow with respect to $N$. We have $\epsilon_{\Delta} \le \epsilon$ if 
{\small
\begin{eqnarray}
r_c \ge 
& \left \{ \noindent
        \begin{array}{ll}
O(N^{4-\alpha \over 4+\alpha}), &  \mbox{$2 \le \alpha <4$,} \\ \nonumber
        &  \nonumber \\ \nonumber
O(\sqrt{\ln N}), & \alpha=4, \nonumber \\ \nonumber
        &  \nonumber \\ \nonumber
O(1), & \alpha>4, \nonumber \\ \nonumber
\end{array} \nonumber
\right.\ \nonumber
\end{eqnarray}
}
\noindent
where $O()$ represents the order for large $N$.}\\  
}

\indent
The proof can be obtained through simple algebraic manipulations from Theorem 1, and is thus omitted\footnote{From Theorem 1, $\mathcal{C}$ is a function of $r_c$; thus, $\mathcal{C}$ is replaced with $r_f^2/r_c^2$.}.\\

\indent
Consider the above expression for large interference where $2 \le \alpha < 4$. 
This corresponds to such channel environments as free space $\alpha=2$, obstructed areas in factories $\alpha \in \{2 \sim 3\}$, 
and urban areas $\alpha \in \{2.7 \sim 3.5\}$. 
\textcolor{black}{$r_c \ge O(N^{4-\alpha \over 4+\alpha})$ shows that $r_c$ grows with $N$ sublinearly.
If communication complexity $\mathcal{C}$ remains constant when network size $N$ increases, 
the maximum number of activated dipoles is 
$o(N^{1-{4-\alpha \over 4+\alpha}})$ = $o(N^{2\alpha \over 4+\alpha})$, 
showing a sparsely activated network. 
Hence, the corresponding Markov Random Field with a small neighborhood size may have insufficiently activated dipoles,  
and may not be an efficient approximation to the global model\footnote{Note that a lower bound would be needed for the approximation error to draw a definitive negative conclusion.}}

Now consider small interference where $\alpha \ge 4$. This corresponds to such channel environments as shadowed urban areas with $\alpha \in \{4 \sim 5\}$, and obstructed regions in buildings ($\alpha \in  \{4 \sim 6\}$) \cite{Rappaport}. When $4 < \alpha \le 6$, $r_c$ $> O(1)$. This implies that $r_c$ can grow with $N$ at an arbitrarily slow rate for the approximation error to be small. The network has densely activated dipoles, i.e., the number of active nodes can be slightly fewer than $O(N)$, showing a densely activated network. This case shows that the Markov Random Field with a small neighborhood now has a sufficient number of activated dipoles,  and is thus an efficient approximation to the global model.\\

\indent
{\bf Topology}: Note that for the uniform network assumed in the theorem, $l_{th}$, the inter-distance between two neighboring nodes, characterizes the physical topology. Thus, Corollary 1 shows that $r_c$ increases with respect to $l_{th}^{1/3}$ for $2 < \alpha \le 4$ and $l_{th}$ for $\alpha=2$. This shows that the rate of growth of the contention range with respect to the inter-node distance.\\

\indent
{\bf Performance-complexity trade-off:} We now examine how $\mathcal{C}$ can vary with respect to $\alpha$ for large $N$.\\ 

\indent
\textcolor{black}{
Corollary 2: {\it Assume $N \gg 1$ (and $\mathcal{C} <N$). $\epsilon_\Delta$ can be simplified as 
{\small
\begin{eqnarray}
\epsilon_{\Delta} &=
\left \{ \noindent
\begin{array}{ll}
{\mathcal{I} \over r_c^2} \left \{ (2+\ln \mathcal{C})^2 + {4\pi \over l_{th}} (\sqrt{Nl_{th} r_c}) +o() \right \}, & \alpha=2 \nonumber \\ \nonumber
&  \nonumber \\ \nonumber
A_4 (N l_{th} r_c)^{4-\alpha \over 4}+o(), & 2< \alpha <4 \nonumber \\ \nonumber
&  \nonumber \\ \nonumber
{\mathcal{I} \over r_c^2} 
\left \{ 
{2(2-{1 \over \mathcal{C}})^2 \over r_c^2} +
{4\pi \over l_{th}^2} ({1 \over \sqrt{\mathcal{C}}} + 
{1 \over 2} \ln({Nl_{th} \over \mathcal{C}r_c})) \right \}, & \alpha=4 \nonumber \\ \nonumber
& \nonumber \\ \nonumber
A_1+A_2 {1 \over \mathcal{C}^{\alpha-2 \over 4}} + A_3 {1 \over \mathcal{C}^{\alpha-4 \over 4}} + O(N^{{4-\alpha}\over 4}), & \mbox{$\alpha >4$}; \nonumber 
\end{array}
\right.\ \nonumber 
\end{eqnarray}
}
\noindent
where $A_1$ = ${{\alpha^2 \mathcal{I}} \over r_c^{\alpha} (\alpha-2)^2}$, $A_2$=${4 \pi \mathcal{I} \over (l_{th}r_c)^{\alpha \over 2}}$, $A_3$=${-2\mathcal{I} \over (4-\alpha)r_c^{\alpha \over 2}}$, $A_4$ = ${\mathcal{I} \over r_c^2}$.\\ 
}
}

\indent
\textcolor{black}{
The proof can be obtained by rewriting $\epsilon_\Delta$ and simple algebraic manipulations, and thus omitted. The corollary shows that the larger the communication complexity $\mathcal{C}$, the larger the Type-1 error, since more dependency is neglected by the local model within an inference range. In addition, a larger $\mathcal{C}$ should result in a smaller Type-2 error. However,  for $ 2\le \alpha \le 4$, the bound for the Type-2 error is nearly independent of $\mathcal{C}$ for large network size $N$. This suggests that the strong interference outside $\mathcal{C}$ dominates for a slow decaying channel, and a Markov Random Field may not efficiently approximate a global model. For $\alpha >4$, $\epsilon_{\Delta}$ monotonically decreases as $\mathcal{C}$ increases, suggesting that the spatial dependence within an interference range dominates for a rapidly decaying channel. In fact, $\epsilon_\Delta$ can be made arbitrarily small if $\mathcal{C}$ and $r_c$ grow at a slow rate with respect to $N$. Markov Random Fields are thus an efficient approximation of the global model.\\
}  

\begin{figure}[htb] \centering
  \begin{tabular}{c}    \resizebox{3.2in}{!}{\includegraphics{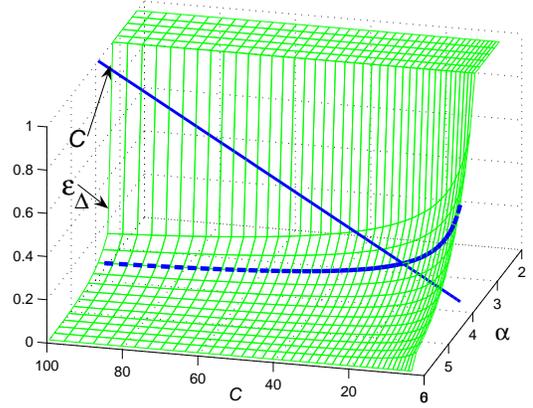}}\\ %alpha_ri_3.eps
   \end{tabular}
\caption{Trade-off between performance and complexity. Thick dashed line: $\alpha$=4.}
\label{tradeoff1}
\end{figure}

\indent
Figure \ref{tradeoff1} shows an example of $\epsilon_\Delta$ as a function of $\mathcal{C}$ and $\alpha$ for $\mbox{SINR}_{th}=20$, $N_b=0.1$, $r_c=10$, $r_f \in \{20 \sim 100\}$, $N=1000$, and $l_{th}=2$ $meter$\footnote{ This corresponds to sensor networks for habitat monitoring, battlefield surveillance, and mechanical measurement and monitoring.}. The flat region in Figure \ref{tradeoff1} corresponds to small $\alpha$ (large interference), where the SINR requirement is violated and $\epsilon_{\Delta}$ is truncated to remain constant for illustration. In contrast, $\epsilon_\Delta$ is less than 10\% for $\alpha=4$ and below 1\% for $\alpha=6$ as shown in the figure.

Figure \ref{tradeoff1} shows a trade-off between the approximation error and the communication complexity for $\alpha=4$. The intersection $\mathcal{C}$ and $\epsilon_{\Delta}$, e.g. between the two thick lines in the figure, corresponds to an optimal neighborhood with $\mathcal{C}$=16. This corresponds to \footnote{In general, an actual optimal value of $r_f$ depends on a constant that weights the relative importance of performance and complexity.} $r_f=40$ and $r_c=10$. In general, for a given $r_f$, $r_c$ can be adjusted to vary $\mathcal{C}$ so that a proper trade-off can be obtained. A wide range of $3$ $\sim$ $20$ hops for $r_c$ would be the most feasible scenarios.\\

%%%%%%%%%%%%%%%%%%
\section{Distributed Algorithm}
\label{SelfManagement}

We now describe randomized distributed algorithms enabled by the local model for self-configuration where nodes make decisions using local information from neighbors. 

%%%%%%%%%%%%%%%%%%%%%
\subsection{Distributed Algorithm}
\label{SO}

The distributed algorithm obtains a near-optimal configuration by maximizing the approximated likelihood function,  
\begin{eqnarray}
(\hat{\mbox{\boldmath$\sigma$}}, \hat{\mbox{\boldmath$X$}}) 
&=& \mbox{arg} \mathop{\mbox{max}}_{\tiny (\mbox{\boldmath$\sigma$}, \mbox{\boldmath$X$})} P^l(\mbox{\boldmath$\sigma_{}$},\mbox{\boldmath$X_{}$}).
\end{eqnarray}

\indent
Due to the spatial Markov property, 
$P^l(\mbox{\boldmath$\sigma_{}$},\mbox{\boldmath$X_{}$})$ is factorizable over cliques. Maximizing the likelihood function reduces to maximizing the local probability density functions (clique potentials), i.e., for $1 \le i,j \le N$, 
\begin{equation}
(\hat{\sigma_{ij}}, \hat{X_i}) 
= \mbox{arg} \mathop{\mbox{max}}_{\tiny (\sigma_{ij}, X_i)} P^l(\sigma_{ij}, X_i | X_{N_i}, \sigma_{N_{ij}}),
\end{equation}
where $P^l(\sigma_{ij}, X_i | X_{N_i}, \sigma_{N_{ij}})$ = $g_{ij}(\mbox{\boldmath$\sigma$}$, $\mbox{\boldmath$X$})$, $X_{N_i}$ and $\sigma_{N_{ij}^I}$ are in the neighborhood of node $i$ and dipole $\sigma_{ij}$, respectively. 
These local probability density functions are composed of the neighboring nodes and dipoles, and the configuration can thus be updated locally. The local maximizations result in coupled equations due to the nested Markov dependence among random variables of $\mbox{\boldmath$\sigma$}$ and  $\mbox{\boldmath$X$}$, showing the need of information exchange among neighbors.

Many algorithms can be used to maximize the local probability density functions, e.g., stochastic relaxation \cite{Geman}\cite{Baras} and message passing \cite{Kschischang}. This work uses stochastic relaxation which is a randomized algorithm. The algorithm converges to the global maximum of $P^l(\mbox{\boldmath$\sigma$},\mbox{\boldmath$X$})$ asymptotically with probability one \cite{Geman}.  

In stochastic relaxation, each node makes local decisions based on a certain probability. 
\textcolor{black}{That is, node $i$ decides $X_i$ and $\{\sigma_{ij}\}$ for $j \in N_i$, which results in a joint optimization of the topology and scheduling.} 
Specifically, let
$\hat{X}_i(t+1)$ and $\hat{\sigma}_{ij}(t+1)$ denote the new position that node $i$ would move to and the activity of link ($i$, $j$) at time $t+1$, respectively. A sequential implementation of the distributed stochastic algorithm at a node $i$ is described as follows: for $t \ge 1$ and $1 \le i \le N$, \\

\indent
(a) given neighbor positions $X_{N_i}$ in the neighborhood $N_i$ of node $i$, determine 
\indent     
$\hat{X}_i(t+1)$ = $x_i$, with probability\\
\indent
$P^l \left (X_i (t+1) = x_i | X_{N_i} (t) \right )= {\exp(-\psi_i(x_i)/T(t+1)) \over \sum_{\forall x_i} \exp(-\psi_i(x_i)/T(t+1))}$, where $\psi_i(x_i)$ =${(x_i - X_{i0})^2 \over 2 \sigma^2}  + \zeta \sum_{j \in N_i^{\theta}} h(x_i,X_j (t))$.\\

\indent
 (b) Given link activities $\sigma_{N_{ij}}$ and positions $ X_{N_i}$ of nodes in the neighborhood of node $i$, \textcolor{black}{determine whether node $i$ transmits to node $j$ for $\forall j \in N_i$}, i.e.,  $\hat{\sigma}_{ij}(t+1)$ = $\sigma_{ij}$ with probability                                           
$P^l \left (\sigma_{ij} (t+1) = \sigma_{ij} | X_{N_i} (t), \sigma_{N_{ij}} (t) \right ) = \\
\indent
{\exp(-\psi_{ij}(\sigma_{ij})/T(t+1)) \over                               \sum_{\forall \sigma_{ij}} \exp(-\psi_{ij}(\sigma_{ij})/T(t+1))}$, where $\psi_{ij}(\sigma_{ij})$ =\\
\indent
$(\alpha_{ij} + \sum_{mn \in N_{ij}^I} [\alpha_{ij,mn} + \alpha_{mn,ij}] {\sigma_{mn}(t)+1 \over 2} )$ ${\sigma_{ij}+1 \over 2}$.\\

\indent
Temperature $T$ is used as a cooling constant in the algorithm, where $T(t)$ = $T_0 / \mbox{log}(1 + t)$ varies with time $t$ and  $T_0$=$3$ \cite{Baras}\cite{MRF}. This allows an almost-sure convergence to the global minimum of the Hamiltonian  (see \cite{Geman} for more details).\\ 
\indent
\textcolor{black}{
Note that the joint configuration results directly from the distributed algorithm, i.e., given $\sigma_{N_{ij}}$ and positions $X_{N_i}$, $(\hat{X}_{i}(t+1), \hat{\sigma}_{ij}(t+1))$ = $(x_i, \sigma_{ij})$ with probability $P^l \left ((X_i(t+1), \sigma_{ij} (t+1)) =(x_i, \sigma_{ij}) | X_{N_i} (t), \sigma_{N_{ij}} (t) \right )$. 
}

%%%%%%%%%%%%%%%%%%%%%%%%%%%%%%%%%
\subsection{Information Exchange}

At time $t$, each node (e.g. node $i$) broadcasts its position and adjacent link status, $(X_i(t)$, $\sigma_{ij}(t)=1)$, to the neighboring active dipoles. At time $t+1$, node $i$ uses the information received from neighbors, i.e., 
$(X_m(t)$, $\sigma_{mn} (t)=1)$ for $m \in N_i$, to update its own local configuration, $X_i(t+1)$ and \{$\sigma_{ij}(t+1)$\}. Here \{$\sigma_{ij}(t+1)$\} denotes the set of adjacent dipoles of node $i$. Meanwhile,  a node that receives a message ($X_i(t)$,  $\sigma_{ij}(t)$=1) also relays the message to its neighbors. 

Such information exchange bears a similar spirit to that of Bellman-Ford routing algorithm. Hence the near-optimal distributed algorithm uses relative geo-location information and link activities from neighbors. Note that as an implementation issue, neighbor positions can also be sensed at a node to avoid information exchange among neighbors.

\section{Self-Configuration: Example and Validation}
\label{SelfConfiguration}

We now apply the distributed algorithm to examples of self-configuration of wireless networks for (a) 1-connectivity formation of physical topology;  
(b) scheduling for a logical configuration; and (c) localized failure adaptation. We also compare the performance of the randomized distributed algorithm with the centralized algorithm from the global model and  the commonly-used distributed protocol model\cite{Gandham}\cite{Gupta}\cite{Luo}.

%%%%%%%%%%%%%%%%%%%%%%%%%%%
\subsection{Example of Self-Configuration}
\label{Example1}

\begin{figure} [htb] \centering
  \begin{tabular}{cc}
    \resizebox{1.5in}{!}{\includegraphics{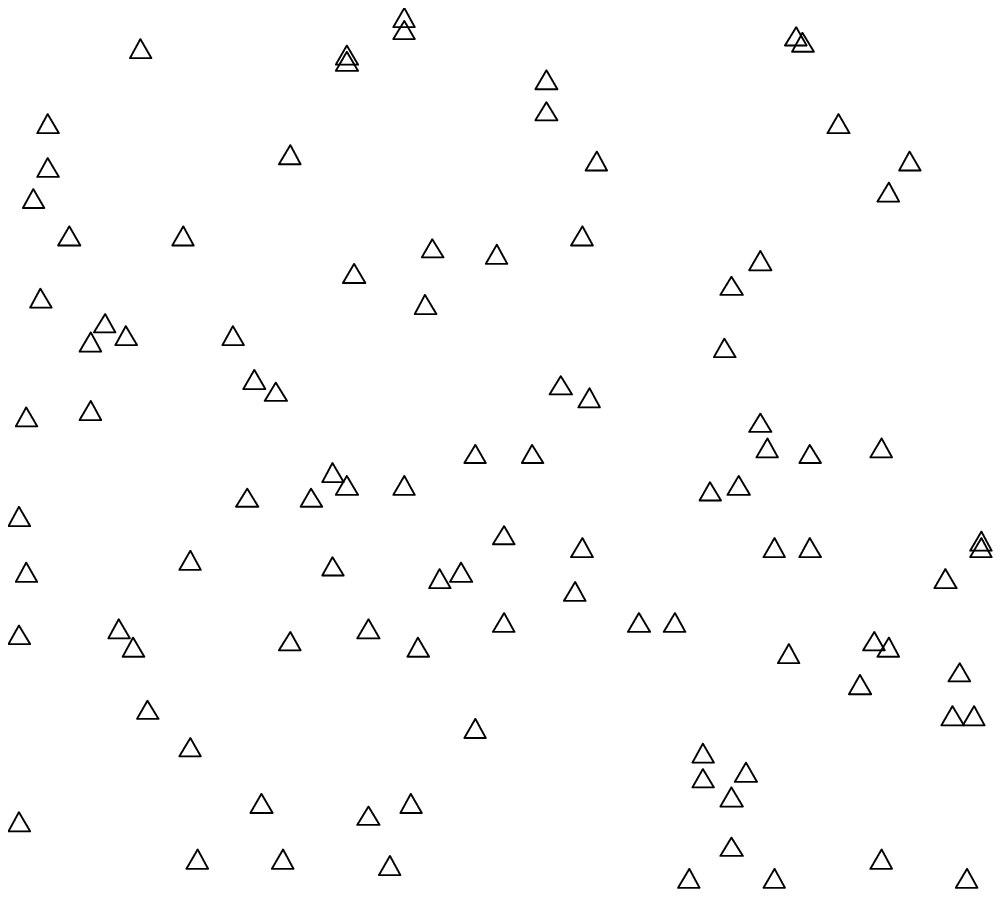}} & 
    \resizebox{1.9in}{!}{\includegraphics{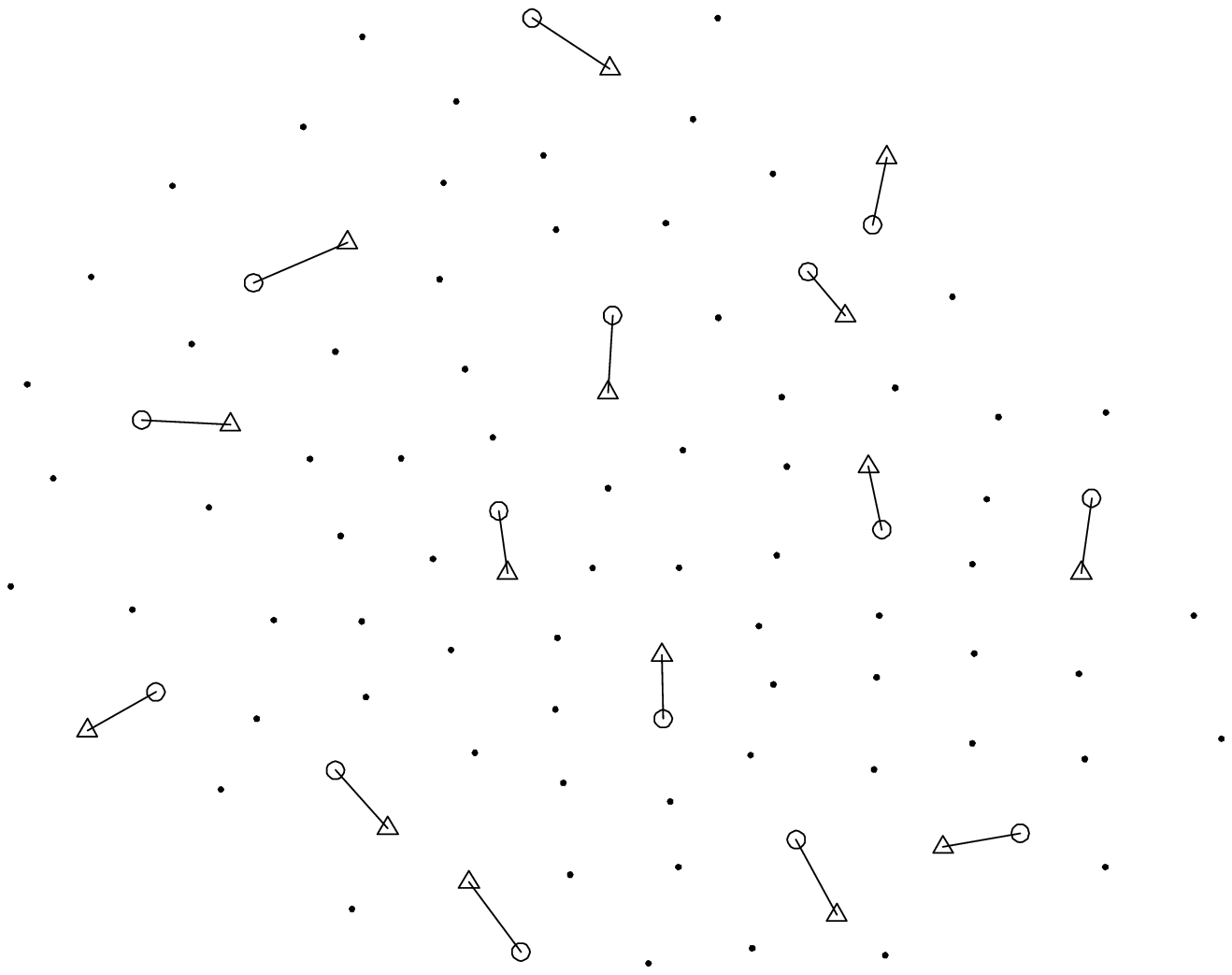}} \\
    {\footnotesize (a) $\mbox{\boldmath$X_0$}$} &    
    {\footnotesize (b) $\mbox{\boldmath$\sigma$}$ given $\mbox{\boldmath$X$}$}
   \end{tabular}
\caption{Self-configuration with localized algorithm}
\label{As}
\end{figure}

We simulate the distributed algorithm for self-configuration using the following parameters:  Network size $N = 100$, the threshold of the inter-node distance  $l_{th} = 2$ and $\theta = {\pi \over 2}$ for the 1-connectivity of the physical topology, $\alpha=4$ for the channel, and $\mbox{SINR}_{th}=20$. Topologies are randomly generated first. Figure \ref{As} (a) shows an example of a random initial configuration ($\mbox{\boldmath$\sigma_0$}$=$\mbox{\boldmath$0$}$, $\mbox{\boldmath$X_0$}$). Each node then updates its position based on the iterative statistical local rules using information from its neighbors as in Section \ref{SO}. The iteration stops when a steady state is reached\footnote{For instance, the condition of $\left |  {|X_i - X_j| - l_{th} \over l_{th}} \right |$ $<$ $0.01$ is satisfied for $\forall i$ and $j \in N_i^{\theta}$.}. 
The dots in Figure \ref{As} (b) show a resulting physical topology with 1-connectivity. The physical topology then remains fixed during formation of a logical configuration. A logical configuration is obtained similarly by the probabilistic local rules in Section \ref{SO} given the physical configuration. Figure \ref{As} (b) also illustrates the resulting logical configuration at the first time epoch. 

\begin{figure} [htb] \centering
\begin{tabular}{cc}
    \resizebox{1.7in}{!}{\includegraphics{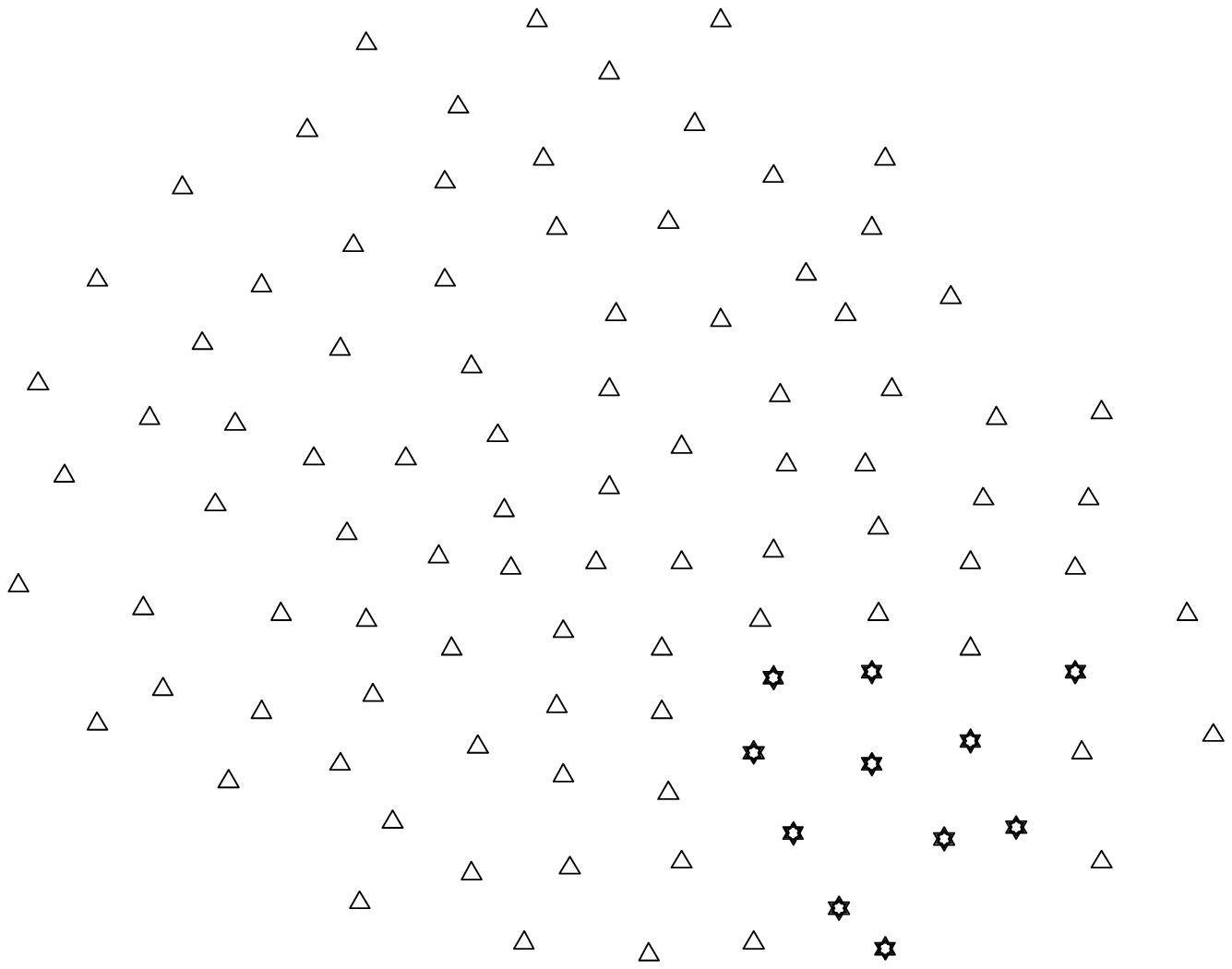}} &
    \resizebox{1.9in}{!}{\includegraphics{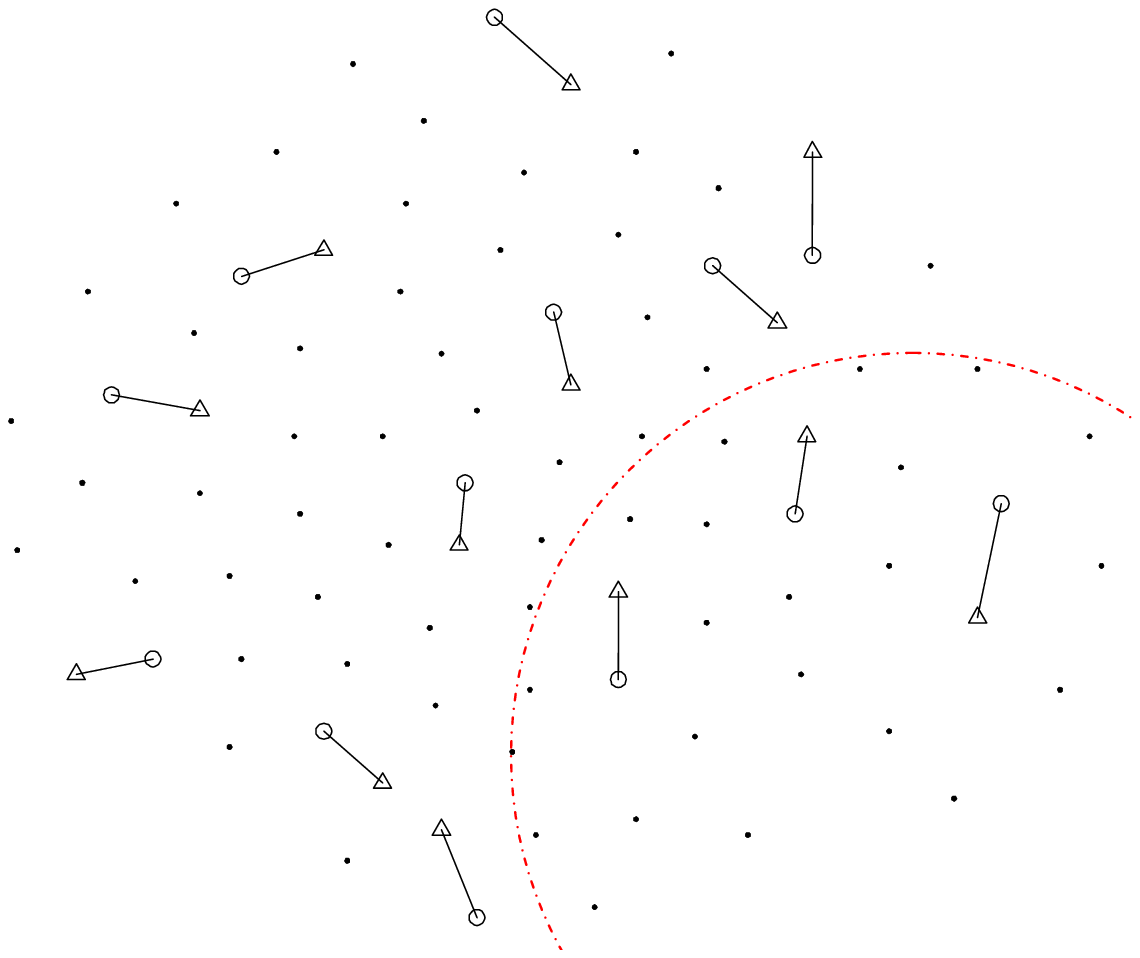}} \\
    {\footnotesize (a) Random failure of nodes } &
    {\footnotesize (b) Localized recovery of $\mbox{\boldmath$\sigma$}$}
   \end{tabular}
\caption{Localized recovery from random node failures}
\label{SR}
\end{figure}

Now, consider that a certain nodes fail in a network configuration shown in Figure \ref{As} (b). Upon failures, the closest neighbors of failed nodes first adjust their positions and select other nodes for transmission. This may cause adjustments to the neighbors and neighbors' neighbors, resulting in cascading changes across the entire network. Hence small perturbations in a network can cause incessant changes to the entire network configuration. Thus localizing failure events is important. Additional penalty terms can be introduced to the Hamiltonian as reconfiguration costs to penalize cascading changes: $\xi \cdot | 
(\mbox{\boldmath$\sigma$},\mbox{\boldmath$X$})-(\mbox{\boldmath$\sigma$}_s,\mbox{\boldmath$X$}_s)|$, where $(\mbox{\boldmath$\sigma$}_s,\mbox{\boldmath$X$}_s)$ denotes the steady-state network configuration, and $\xi$ is a positive weighting constant that characterizes the cost of change in node positions and/or link activities. Such a constraint would localize the change.

Figure \ref{SR} (a) shows failures of wireless nodes that are marked as stars. Localized recovery of the physical topology is shown in Figure \ref{SR} (b). The nodes outside the arc are not affected by failures; the resulting configuration, however, is no longer globally optimal. The failed logical topology can be locally recovered using the same algorithm also shown in Figure \ref{SR} (b).

\begin{figure} [htb] \centering
  \begin{tabular}{c}
   \resizebox{2.2in}{!}{\includegraphics{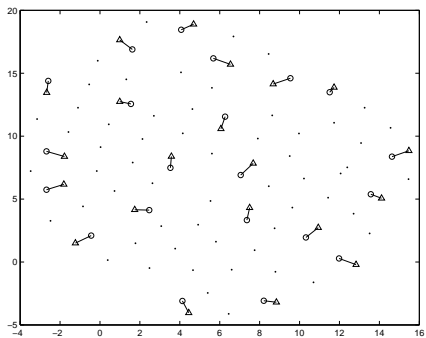}} 
   \end{tabular}
\caption{\textcolor{black}{Joint Optimization of Link Activities and Node Positions}}
\label{JointCoupled}
\end{figure}

\indent
\textcolor{black}{
Figure \ref{JointCoupled} shows the joint optimization of $\mbox{\boldmath$\sigma$}$ and $\mbox{\boldmath$X$}$. The joint optimization is done at a node-dipole pair, i.e., $\{X_i, \sigma_{ij}\}$ for $j \in N_i$ first, and then moved on to another node-dipole pair $\{X_m, \sigma_{mn}\}$ for $n \in N_m$, and so on. Compared to the sequential configuration in Figure \ref{As}.(b), the joint optimization in Figure \ref{JointCoupled} increases the number of active dipoles by 30\% (from $15$ to $20$).
}

\begin{figure} [htb] \centering
  \begin{tabular}{c}
   \resizebox{2.7in}{!}{\includegraphics{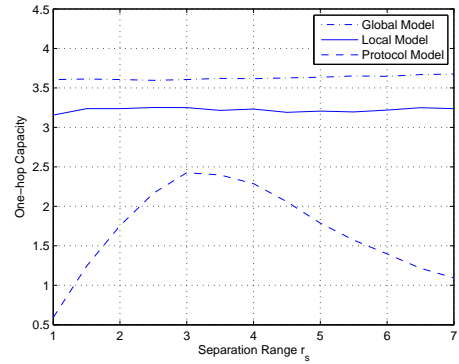}} 
   \end{tabular}
\caption{Comparison of One-hop Capacity: Global, Local, and Protocol Model}
\label{Net_One_Hop}
\end{figure}

%%%%%%%%%%%%%%%%%%%%%%%%%%%%%%%%%
\subsection{Model Validation}

We now validate the performance of the local model, communication complexity and the trade-offs.\\ 

\indent
{\bf Performance:} We first examine through simulation the performance of  a local model compared with that of the global model and commonly-used protocol models in the prior work \cite{Gandham}\cite{Gupta}\cite{Luo}. 

A wireless network is generated with $50$ randomly-positioned nodes,  $\alpha=4$, and $\mbox{SINR}_{th}=20$.  10 simulations are conducted  using random initial topologies and link activities, and the results are averaged. The interference range is the entire network for the global model, and chosen to be $r_f$=$4$ for the local model. The protocol model assumes a separation (i.e., contention) range of $r_s$ without using an interference range. 

Figure \ref{Net_One_Hop} shows the one-hop capacity achieved by the global model, the local model, and the protocol model, respectively. For $r_s$ large, the protocol model fails to maximize the spatial channel-reuse. This is because the contention constraint is too stringent. For $r_s$ small, the protocol model over-utilizes the channel resource, resulting in a violation of SINR constraint and a reduced spatial-reuse. Even at the optimal separation range ($r_s$=$3$) the spatial-reuse of the protocol model is less than that of the local model by $23\%$. \textcolor{black}{The inefficient spatial-reuse is due to the protocol model that cannot take a non-circular contention region  into consideration}. On the contrary, the global and local models characterize the interference and non-circular contention regions and thus represent the neighborhood system of dipoles more accurately. Compared to the global model, the local model has $8\%$ less spatial reuse on the average, showing the performance degradation of a simpler model.

\begin{figure} [htb] \centering
  \begin{tabular}{c}   \resizebox{2.5in}{!}{\includegraphics{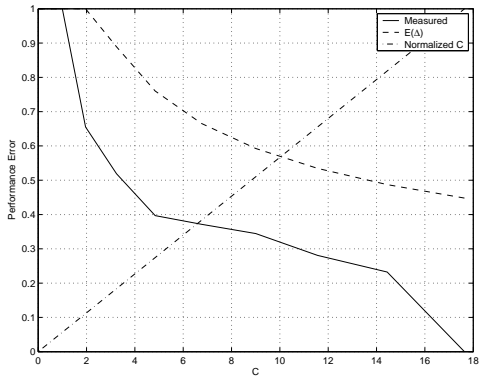}} %CC.eps
   \end{tabular}
\caption{\textcolor{black}{Measured Approximation Error and Upper Bound $E(\Delta)$}}
\label{E_Delta_Comparison}
\end{figure}

\textcolor{black}{
We now examine the tightness of the bound on the approximation error. Specifically, the approximation error is measured from simulations and compared with that calculated from Theorem 1. A linear topology is selected, where 100 nodes are randomly placed and a node communicates with two neighbors. A linear topology is chosen as it provides a worst case of non-circular contention- and interference-regions. 
Other parameters used in the simulations are $\alpha=4$, $\mbox{SINR}_{th}$=10, $N$=100, $r_c$=10, and a varying $r_f$ to obtain values for $\mathcal{C}$. Results are shown in Figure \ref{E_Delta_Comparison}. The measured approximation error decreases sharply as $\mathcal{C}$ increases and is indeed bounded by $E(\Delta)$. The bound follows the same trend as the actual approximation error. The difference between the measured approximation error and the bound may result from the linear topology whose density of active links is much sparser than the assumptions in Theorem 1. Intersections between $C$ and $E[\Delta]$ (or the bound) illustrate the performance-complexity trade-off.
}

\begin{figure} [htb] \centering
  \begin{tabular}{c}
   \resizebox{2.8in}{!}{\includegraphics{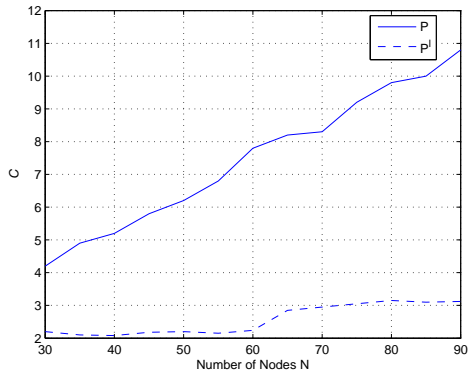}} %CC.eps
   \end{tabular}
\caption{Communication complexity $\mathcal{C}$ vs network size $N$. $P$ and $P^l$: global and local model, respectively.}
\label{Cap1}
\end{figure}

\indent
{\bf Complexity:} 
We now compare the communication complexity of a local and a global model through simulations. We use the same parameters and the number of runs in the simulation. The actual communication complexity is obtained by counting the number of the neighboring active links within the interference range of each active dipole, and averaged over all active dipoles and 10 runs. Figure \ref{Cap1} shows the communication complexity for both the global and local model as a function of network size $N$. The communication complexity of the centralized global optimization increases linearly with $N$ since each node uses the information from all other nodes in the network. The complexity increases slowly with $N$ for the distributed algorithm.\\

%Sung-eok: Change the y-label of Fig. 12 to C.  

%
\vspace{-0.1cm}
\begin{figure} [htb] \centering
  \begin{tabular}{c}
%\resizebox{2.7in}{!}{\includegraphics{Cycle_.eps}}
%\\
%{\footnotesize (a) Timeslot Allocation in a Random Topology}\\
\resizebox{2.5in}{!}{\includegraphics{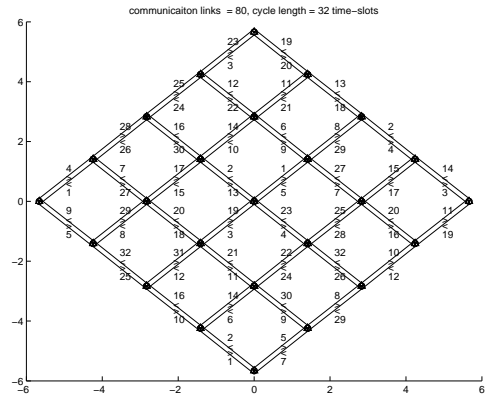}}\\
%{\footnotesize (b) Timeslot Allocation in a Lattice Topology}
   \end{tabular}
\caption{\textcolor{black}{Timeslot Allocation of Spatial Time Division Multiple Access (STDMA)}}
\label{Cycle}
\end{figure}

\textcolor{black}{{\bf Extension}: We now consider an extension to multiple time-slots to show the promise of the distributed algorithm to scheduling in a more realistic setting. This is motivated by the prior work \cite{Modiano} that shows optimal scheduling decisions can be done in a fully distributed fashion. \cite{Modiano} considers packet-level scheduling, the queue size and packet arrivals but a simplified interference model. Our model does not include packet arrivals and queue occupancy but uses a general interference model and adapts physical topology also.  To compare these two works at a common ground, we assume that there is one queue for each communication link and a queue is always busy, and a physical topology is given. Our local probabilistic model is extended directly to include multiple time-slots in the coefficients and details are given in Appendix \ref{Model_STDMA}. We conduct a simulation for scheduling using the following parameters: $\alpha=4$, $N$= 25, $l_{th}$=2, and $r_f$=6 for regular topology in Figure \ref{Cycle}. There are 80 links and 32 assigned time-slots. The figure shows that the optimal allocations are obtained in these examples by the local model and the distributed algorithm while satisfying the SINR constraints. }
\textcolor{black}{
Hence, compared with \cite{Modiano} when a physical topology is given, an advantage of the local model is its ability to approximate the general (non-Markovian) SINR model. A disadvantage of our local model is a lack of specifications of network traffic.} 

%Removed some redundancies. 
	
%%%%%%%%%%%%%%%%%%%%%%%%%%%%%%%%%%%%%%%%
\section{Conclusion}
\label{Conclusions}

We have developed an analytical framework in which near-optimality, approximation and randomized distributed algorithms can be studied for self-configuration, i.e., joint configuration of a topology and scheduling. Our findings are summarized as follows.

(a) We begin with a global model that characterizes the ground truth in regard to network assumptions. The ground truth includes the randomness from a network internally and management constraints imposed externally.  The resulting model is a Gibbs distribution where the exponent can be regarded as a cost function. This relates modeling with optimization so that both randomness in a network and objectives can be combined naturally. This approach differs from the pure ``emerging behavior" where nodal decisions are not governed by an optimal performance. This approach also differs from the ``black-box" method where models are learned externally.

A disadvantage of our approach is the simple assumptions that limit the current model. Fading has not been included in our model. Traffic demands at network-layer have been considered in developing a probabilistic graphical model in \cite{Liu05} but not yet in this work. 

\indent
(b) The mathematical representation of the global model quantifies the statistical spatial dependence of node positions and node-node communication. The complex spatial dependence is represented explicitly by a probabilistic graph. The graph shows how to obtain a local model to approximate the global model. The local model is a two-layer Markov Random Field or a random bond model. The complexity of the local model is  the communication range of nodes from neighborhoods in the Markov Random Field. We remove long links in the graph to obtain a local model, and a more sophisticated approach may be explored to obtain a better approximation. 

\indent
(c) We have derived sufficient conditions on the near-optimality of the local model under different channel conditions, density of nodes, network size and complexity. \textcolor{black}{The conditions show a trade-off between the near-optimality and complexity. For example, a local model is near-optimal with a moderate complexity if a channel has a fast power decay of order at least $4$; and may not be near-optimal, otherwise. The near-optimality conditions thus complement the empirical choices in protocol designs (e.g. 802.11), and may provide a practical utility in regard to the effect of channel attenuation, densities of nodes, and node-pair communications.} 

\indent
(d) Near-optimal local models render a class of randomized distributed algorithms for self-configuration. The algorithm allows each node to adapt probabilistically its local configuration using only information from neighbors. Local self-configuration collectively achieves a near-optimal global configuration. Also, the distributed algorithm achieves a near-optimal configuration at a bounded communication complexity. Hence the algorithmic advantage is a strength of the local model. We have shown examples of stochastic scheduling and reconfiguration upon failures, and compared the performance and complexity with existing protocol models. 

One disadvantage of stochastic relaxation is the slow convergence that has been explored in the prior work \cite{hajek88}. Other simple randomized algorithms have been discussed in \cite{lin06}. It would be beneficial to study simpler randomized algorithms.
%

%%%%%%%%%%%%%%%%%%% APPENDIX %%%%%%%%%%%%%%%%%%%%%%%%%%%%%%%%%%%%%%%%

\section*{Appendix}
\setcounter{section}{0}
\section{Proof of Theorem 1}
\label{Theorem1}

{\small
Proof: 
We begin proving the theorem by bounding the error, i.e.,      $|{H(\mbox{\boldmath$\sigma_{}$}^*|{\mbox{\boldmath$X_{}$}}) - 
   H(\hat{\mbox{\boldmath$\sigma_{}$}}|{\mbox{\boldmath$X_{}$}}) \over H({\mbox{\boldmath$\sigma_{}$}}^*|{\mbox{\boldmath$X_{}$}})}| 
\le  
{|H(\mbox{\boldmath$\sigma_{}$}^*|{\mbox{\boldmath$X_{}$}}) - H(\hat{\mbox{\boldmath$\sigma_{}$}}|{\mbox{\boldmath$X_{}$}})|^{u}
\over |H({\mbox{\boldmath$\sigma_{}$}}^*|{\mbox{\boldmath$X_{}$}})|^{l} },$
}
\noindent
where the super-scripts $u$ and $l$ denote an upper and a lower bound of the corresponding quantity.

\textcolor{black}{
{\small
To find an upper bound of the numerator, we have  $|H(\mbox{\boldmath$\sigma_{}$}^*|{\mbox{\boldmath$X_{}$}}) - H(\hat{\mbox{\boldmath$\sigma_{}$}}|{\mbox{\boldmath$X_{}$}})|$ 
$=$
$|(H(\mbox{\boldmath$\sigma_{}$}^*|{\mbox{\boldmath$X_{}$}}) - H^l({\mbox{\boldmath$\sigma_{}$}^*}|{\mbox{\boldmath$X_{}$}}))+ 
(H^l(\hat{\mbox{\boldmath$\sigma_{}$}}|{\mbox{\boldmath$X_{}$}})) - 
H(\hat{\mbox{\boldmath$\sigma_{}$}}|{\mbox{\boldmath$X_{}$}}) +
(H^l({\mbox{\boldmath$\sigma_{}$}^*}|{\mbox{\boldmath$X_{}$}})-H^l(\hat{\mbox{\boldmath$\sigma_{}$}}|{\mbox{\boldmath$X_{}$}}))|$
$\le$  
$|H(\mbox{\boldmath$\sigma_{}$}^*|{\mbox{\boldmath$X_{}$}}) - H^l({\mbox{\boldmath$\sigma_{}$}^*}|{\mbox{\boldmath$X_{}$}})| +
|H(\hat{\mbox{\boldmath$\sigma_{}$}}|{\mbox{\boldmath$X_{}$}}) - H^l(\hat{\mbox{\boldmath$\sigma_{}$}}|{\mbox{\boldmath$X_{}$}})|$,
}
where the inequality holds since 
{\small $H(\mbox{\boldmath$\sigma_{}$}^*|{\mbox{\boldmath$X_{}$}})$ $\le$
$H(\hat{\mbox{\boldmath$\sigma_{}$}}|{\mbox{\boldmath$X_{}$}})$, and 
$H^l(\hat{\mbox{\boldmath$\sigma_{}$}}|{\mbox{\boldmath$X_{}$}})$ $\le$  
$H^l({\mbox{\boldmath$\sigma$}^*}|{\mbox{\boldmath$X_{}$}})$ by definition. 
}
}

{\small
For any configuration ($\mbox{\boldmath$\sigma$}|\mbox{\boldmath$X$}$),
$|H(\mbox{\boldmath$\sigma$}|\mbox{\boldmath$X$})     - H^l(\mbox{\boldmath$\sigma$}|\mbox{\boldmath$X$})|$ 
$\le$
\textcolor{black}{
$|R_I(\mbox{\boldmath$\sigma$}|\mbox{\boldmath$X$})|$+ $|R_3(\mbox{\boldmath$\sigma$}|\mbox{\boldmath$X$})|$}. 
}
Let $I_3$ and $I_R$ be an upper bound of 
\textcolor{black}{$|R_3(\mbox{\boldmath$\sigma$}|\mbox{\boldmath$X$})|$ and $|R_I(\mbox{\boldmath$\sigma$}|\mbox{\boldmath$X$})|$, respectively, 
}
where

{\small
\begin{eqnarray}
\label{R_3}
|R_3({\mbox{\boldmath$\sigma$}}|{\mbox{\boldmath$X$}})| &\le& 
\sum_{k_1=1}^{ r_f \over r_c } \sum_{k_2=1}^{ r_f \over r_c }
    2 P_{t} k_1^{{-\alpha \over 2}} k_2^{{-\alpha \over 2}} r_c^{-\alpha}
\nonumber \\ \nonumber
&\le& 2 P_{t} \left (1 + (\int_{k=1}^{ r_f \over r_c } k^{{-\alpha \over 2}} dk) \right )^2 \cdot r_c^{-\alpha}
\nonumber \\ \nonumber
&=& \left \{
        \begin{array}{ll}
        2P_{t} r_c^{-\alpha} (1 + 0.5 \ln{\mathcal{C}})^2, 
        &  \alpha=2 \nonumber \\ \nonumber
        &  \nonumber \\ \nonumber
        {2P_{t} r_c^{-\alpha} \over (\alpha -2)^2}  
(\alpha - 2({\mathcal{C}})^{{2-\alpha \over 2}})^2, 
        &  \alpha > 2 \nonumber
        \end{array} \nonumber
\right.\ \nonumber \\ \nonumber
& & \nonumber \\ \nonumber
&=& I_{3},  \nonumber
\end{eqnarray}
}
\noindent
\textcolor{black}{
{\small where ``1" appears in the above expression to ensure that the discrete sum upper bounds the integral. The penalty term in the configuration Hamiltonian is not included as $I_3$ is an upper bound. \\
\indent
We now find $I_R$. Consider an active dipole $\sigma_{ij}$, and let $G_k$ be the set of neighboring active dipoles that are $r_f+kr_c$ apart from the receiver. The cardinality of $G_k$ is upper bounded, i.e., 
$|G_k|$ $\le$ ${2 \pi \over \theta}$ = $\pi / \sin^{-1} \left ({ {r_c \over 2} \over r_f + (k-1) r_c} \right )$ $<$ $2 \pi (r_f + (k-1)r_c) / r_c$. 
}
}

{\small For an active dipole $\sigma_{ij}$, let $I_{kU}$ be an upper bound of the residual interference outside interference range\footnote{Similar to $I_3$, the penalty term is not included for deriving an upper bound.}, i.e.,
}
\textcolor{black}{
{\small
\begin{eqnarray}
\label{I_k_U}
R_{I_{ij}}(\mbox{\boldmath$\sigma$}|\mbox{\boldmath$X$}) 
&\le& \sum_{k=1}^{k_U} 2 P_{t} 2 \pi \cdot {(r_f + (k-1)r_c)^{2-\alpha \over 2} \over r_c} (l_{th} ( 1 +\epsilon_0))^{-\alpha \over 2} \nonumber \\ \nonumber
&=& I_{k_U}, \nonumber
\end{eqnarray}
}
}
\noindent
where $k_U$ is an integer that satisfies the inequality  
{\small
\begin{equation}
\label{Quadratic}
{(N-2) l_{th} (1 + \epsilon_0) \over r_c} \le \sum_{k=1}^{k_U} 2 \pi { r_f + (k-1)r_c \over r_c}.  \end{equation}
}
\noindent

{\normalsize The value ${(N-2)l_{th} (1 + \epsilon_0) \over r_c}$ denotes an upper bound of the maximum number of available active dipoles except $\sigma_{ij}$ (with two nodes) in a network with total $N$ nodes.  
$k_U$ can be solved from the above inequality as 
$k_U \le r_c+ \sqrt{N l_{th} r_c}$. Using this bound for $k_U$ and replacing $r_f$ by $\sqrt{\mathcal{C}}r_c$, we have 
}
{\small
\begin{eqnarray}
\label{I_R}
& & I_{k_U} \nonumber \\ \nonumber
& & \nonumber \\ \nonumber
 &\le& {2 P_{t} 2 \pi \over r_c (l_{th} ( 1 +\epsilon_0))^{\alpha \over 2} } \left ( r_f^{2-\alpha \over 2} + \int_{1}^{k_U^u} (r_f + (k-1)r_c)^{2-\alpha \over 2} dk \right ) \nonumber \\ \nonumber
 & & \nonumber \\ \nonumber
 &\le& \left \{ \noindent
\begin{array}{ll}
{P_{t}4\pi \over r_c^2 (l_{th}(1 +\epsilon_0))^{\alpha \over 2}} \left ( {1 \over \sqrt{\mathcal{C}}} + \ln (1+\sqrt{Nl_{th} /(\mathcal{C}r_c) } \right ), & \alpha=4 \nonumber \\ \nonumber
&  \nonumber \\ \nonumber
{P_{t}4\pi \over r_c^2 (l_{th} (1 +\epsilon_0))^{\alpha \over 2}}  \left ( \mathcal{C}^{2-\alpha \over 4} r_c^{4-\alpha \over 2} + {2 \over 4-\alpha} \cdot  \right . \nonumber \\ \nonumber   
& \nonumber  \\ \nonumber
\left . \left (-(\sqrt{\mathcal{C}}r_c)^{4-\alpha \over 2} + (\sqrt{\mathcal{C}}r_c + \sqrt{Nl_{th} r_c})^{4-\alpha \over 2} \right ) \right ), & \alpha \neq 4. \nonumber 
\end{array}
\right.\ \nonumber \\ \nonumber
& & \nonumber \\ \nonumber
&=& I_R. \nonumber
\end{eqnarray}
}
\indent
Thus,
{\small
$|H(\mbox{\boldmath$\sigma_{}$}^*|{\mbox{\boldmath$X_{}$}})$ $-$ $H(\hat{\mbox{\boldmath$\sigma_{}$}}|{\mbox{\boldmath$X_{}$}})|$
$\le$ 
$|H(\hat{\mbox{\boldmath$\sigma_{}$}}|{\mbox{\boldmath$X_{}$}})$ $-$ 
$H^l(\hat{\mbox{\boldmath$\sigma_{}$}}|{\mbox{\boldmath$X_{}$}})|$ 
$+$
$|H({\mbox{\boldmath$\sigma_{}$}}^*|{\mbox{\boldmath$X_{}$}})$ $-$ 
$H^l({\mbox{\boldmath$\sigma_{}$}}^*|{\mbox{\boldmath$X_{}$}})|$
$\le$ $2 (I_R + I_3) {N_{\sigma}}^*$,
}
where ${N_{\sigma}}^*$ is total number of active dipoles in ${\mbox{\boldmath$\sigma$}}^*$ given ${\mbox{\boldmath$X$}}$.\\
\indent
{\normalsize
To obtain a lower bound for $|H({\mbox{\boldmath$\sigma_{}$}}^*|{\mbox{\boldmath$X_{}$}})|$, we 
take only the net energy term in $ H({\mbox{\boldmath$\sigma_{}$}}^*|{\mbox{\boldmath$X_{}$}})$ as the penalty term is usually small, i.e., close to zero when the constraint is satisfied. Then    
}
{\small $|H({\mbox{\boldmath$\sigma_{}$}}^*|{\mbox{\boldmath$X_{}$}})|$ 
$\ge$ $\min\{ P_{t} l_{ij}^{-\alpha} -$  $\sum_{mn \neq ij} P_{t} l_{mj}^{-\alpha \over 2}$ $l_{ij}^{-\alpha \over 2} \}$  $N_{\sigma}^*$ for $\forall$ 
$\sigma_{ij}=1$}. 
For $\sigma_{ij}=1$, to satisfy a given $\mbox{SINR}_{th}$, a sufficient condition is {\small $(P_{t} (\sum_{mn \neq ij} l_{mj}^{-\alpha \over 2})^2 + N_b) / (P_{t} (l_{ij}^{-\alpha \over 2})^2)$ $\le$ $1/\mbox{SINR}_{th}$, thus $(\sum_{mn \neq ij} l_{mj}^{-\alpha \over 2})^2$ $\le$ $({l_{ij}^{-\alpha} \over \mbox{SINR}_{th}} - {{N_b} \over P_{t}})$}. 
Therefore, 
{\small $|H({\mbox{\boldmath$\sigma_{}$}}^*|{\mbox{\boldmath$X_{}$}})|$ $\ge$
($P_{t} l_{th}^{-\alpha}(l+\epsilon_0)^{-\alpha}$ $-$
$P_{t} l_{th}^{-\alpha \over 2}(l-\epsilon_0)^{-\alpha \over 2}$ 
$\sqrt{ l_{th}^{-\alpha \over 2} (1-\epsilon_0)^{-\alpha \over 2}  \mbox{SINR}_{th}^{-1} - {N_b \over P_{t}}}$) $N_{\sigma}^*$.\\
}

\indent
Hence,
{\small
$|{H(\mbox{\boldmath$\sigma_{}$}^*|{\mbox{\boldmath$X_{}$}}) - 
   H(\hat{\mbox{\boldmath$\sigma_{}$}}|{\mbox{\boldmath$X_{}$}}) \over 
   H({\mbox{\boldmath$\sigma_{}$}}^*|{\mbox{\boldmath$X_{}$}})}|]$ 
$\le$ $\epsilon_{\Delta}$, where 
$\epsilon_{\Delta}$ = $2 (I_R + I_3) / I_D$, $I_D$ = $P_{t} l_{th}^{-\alpha}(l+\epsilon_0)^{-\alpha} - P_{t} l_{th}^{-\alpha \over 2}(l-\epsilon_0)^{-\alpha \over 2} \sqrt{ l_{th}^{-\alpha \over 2} (1-\epsilon_0)^{-\alpha \over 2} / \mbox{SINR}_{th}-{N_b \over P_{t}}}$.\\
}

\noindent
Assume $\epsilon_0$=0 for a simple representation,
{\small
\begin{eqnarray}
\epsilon_{\Delta} &=
\left \{ \noindent
\begin{array}{ll}
{\mathcal{I} \over r_c^2} \left ( (2+\ln \mathcal{C})^2 + {4\pi \over l_{th}} (r_c+\sqrt{Nl_{th} r_c}) \right ), & \alpha=2 \nonumber \\ \nonumber
&  \nonumber \\ \nonumber
{\mathcal{I} \over r_c^2} \left ( {2 (2-{1 \over \mathcal{C}})^2 \over r_c^2}+{4\pi \over l_{th}^2}({1 \over \sqrt{\mathcal{C}}}+\ln(1+\sqrt{{Nl_{th} \over \mathcal{C}r_c}})) \right ), & \alpha=4 \nonumber \\ \nonumber
& \nonumber \\ \nonumber
{\mathcal{I} \over r_c^2} \left ( {2(\alpha-2\mathcal{C}^{2-\alpha \over 2})^2  \over (\alpha-2)^2 r_c^{\alpha-2} }  + {4\pi \over l_{th}^{\alpha \over 2}} \left (\mathcal{C}^{2-\alpha \over 4} r_c^{4-\alpha \over 2} +{2 \over 4-\alpha} \left ( \right. \right. \right .& \nonumber \\ \nonumber
\left . \left. \left.-(\sqrt{\mathcal{C}}r_c)^{4-\alpha \over 2} +(\sqrt{\mathcal{C}}r_c+\sqrt{Nl_{th}r_c})^{4-\alpha \over 2} \right ) \right ) \right ), & \mbox{o.w.}, \nonumber
\end{array}
\right.\ \nonumber 
\end{eqnarray}
\noindent
where $\mathcal{I}$= $2 / \left(l_{th}^{-\alpha \over 2} ( l_{th}^{-\alpha \over 2}-\sqrt{ l_{th}^{-\alpha}/\mbox{SINR}_{th}-N_b / P_{t}} ) \right)$.
}

\indent
\textcolor{black}{
Moreover, since the bounds derived above hold for any $(\sigma, X)$, they also holds for the expected value, i.e., 
{\small
$E[|{H(\mbox{\boldmath$\sigma_{}$}^*,{\mbox{\boldmath$X_{}$}}^*) - H(\hat{\mbox{\boldmath$\sigma_{}$}},\hat{\mbox{\boldmath$X_{}$}}) \over
   H(\mbox{\boldmath$\sigma_{}$}^*,{\mbox{\boldmath$X_{}$}}^*)}|] \le \epsilon_\Delta$.
}
}

\section{Model Extension to Multiple Timeslots}
\label{Model_STDMA}

\textcolor{black}{
{\small
As the slot allocation is based on the spatial dependence among neighboring nodes and links, we extend the local model and the distributed algorithm to spatial time division multiple access (STDMA). We assign a time slot $S_{ij}$ to a dipole ($i,j$), $1$ $\le$ $S_{ij}$ $\le$ $S_{max}$, where $S_{max}$ is the TDMA cycle length. Since $S_{max}$ is unknown, we then generalize the coefficients in the Hamiltonian in (\ref{H1}) as functions of time slots,  
}
}
\textcolor{black}{%
{\small
\begin{eqnarray}
\label{Alpha2}
\alpha_{ij}^{'}     &=& \alpha_{ij} \cdot \omega_s (S_{ij}),  \\ 
\alpha_{ij,mn}^{'}  &=& \alpha_{ij,mn} \delta(S_{ij},S_{mn}) \cdot \omega_s(S_{ij}) \nonumber \\ \nonumber
\alpha_{ij,mn,uv}^{'}  &=& \alpha_{ij,mn,uv} \delta(S_{ij},S_{mn}) \delta(S_{ij},S_{uv}) \cdot \omega_s(S_{ij}), \nonumber
\end{eqnarray}
}
\noindent
{\small
where $S_{ij}$ is the time-slot number of TDMA cycle for dipole ($i,j$), and $\delta(S_{ij},S_{mn})$=1 if $S_{ij}=S_{mn}$; $\delta(S_{ij},S_{mn})$=0, otherwise. $\omega_s(S_{ij})$ is a function that is inverse-proportional to $S_{ij}$, and $\omega_s(S_{ij})$ = ${1 \over S_{ij}}$ is used in this work. This enables early sequences of  timeslots to play an more important role in minimizing the potential energy. For example, the first timeslot is utilized first and so on. As a result,  the length of STDMA cycle $S_{max}$ is minimized. 
}
}

\textcolor{black}{
{\small The resulting MRF model is now defined over multiple orthogonal time slots, i.e. $S_{max}$ time slots, and a communication dipole $\sigma_{ij}$ now takes an extended form of $\sigma_{ij}$ $\cdot$ $\delta$$(\mbox{\boldmath$S$}_{ij}-s)$, where $\delta$$( \mbox{\boldmath$S$}_{ij}-s)$=1 if $\mbox{\boldmath$S$}_{ij}$=$s$; 0, otherwise, for $1 \le s \le S_{max}$.\\
}
}

\section*{Acknowledgement} The authors gratefully acknowledge the support from NSF ECS 0300605 and ECS 990857. The authors would like to thank  anonymous reviewers for helpful comments. C.Ji would like to thank C.S. Ji and J. Modestino for helpful discussions on Markov Random Fields.

%%%%%%%%%%%%%%%%%%% REFERENCES %%%%%%%%%%%%%%%%%%%%%%%%%%%%%%%%%%%%%%%%
\nocite{*}
\bibliographystyle{paper}

\vspace{-2cm}
\begin{biography}{Sung-eok Jeon} received the B.S. degree from Yonsei University, Seoul, Korea, in 1996, the M.S. degree from KAIST, Daejeon, Korea, in 1999, the Ph.D degree from Georgia Institute of Technology, in 2007, all in Electrical Engineering. He is now in Microsoft. His research interests are in statistical distributed management of large and complex networks, based on the probabilistic graphical models.
\end{biography}

\vspace{-2cm}
\begin{biography}{Chuanyi Ji}'s research lies in the areas of networking, 
machine learning, and their interfaces. Her research interests are in understanding and managing complex networks, applications of machine learning to network
management and security, large-scale measurements, learning theory and algorithms, and information theory. Chuanyi Ji received the B.S. (Honors)
degree from Tsinghua University, Beijing,
China, in 1983, the M.S. degree from the University
of Pennsylvania, Philadelphia, in 1986, and
the Ph.D. degree from the California Institute of
Technology, Pasadena, in 1992, all in electrical
engineering. She is an Associate Professor in the
Department of Electrical and Computer Engineering,
Georgia Institute of Technology, Atlanta. She was on
the faculty at Rensselaer Polytechnic Institute, Troy,
NY, from 1991 to 2001. Chuanyi Ji received an NSF Career Award in 1995, and an Early Career Award from
Rensselaer Polytechnic Institute in 2000.
\end{biography}

\end{document}